\documentclass[twocolumn]{aastex63}

\usepackage{graphicx,subfigure}
\usepackage[utf8]{inputenc}
\usepackage{longtable}
\usepackage{booktabs}
\usepackage{multirow}
\usepackage{savesym}
\savesymbol{tablenum}
\usepackage{siunitx}
\restoresymbol{SIX}{tablenum}
\usepackage{natbib}
\usepackage{amsmath}
\graphicspath{{./}{figures/}}
\accepted{for publication in ApJ}

\begin{document}

\title{X-ray spectroscopy in the microcalorimeter era III: line formation under Case A, Case B, Case C, and Case D in H- and He-like iron for a photoionized cloud}
\author[0000-0002-4469-2518]{P Chakraborty}
\affiliation{University of Kentucky \\
Lexington, KY, USA}
\author[0000-0003-4503-6333]{G. J. Ferland}
\affiliation{University of Kentucky \\
Lexington, KY, USA}
\author[0000-0002-8823-0606]{M. Chatzikos}
\affiliation{University of Kentucky \\
Lexington, KY, USA}
\author[0000-0002-2915-3612]{F. Guzm\'an}
\affiliation{University of North Georgia \\
Dahlonega, GA, USA}
\author[0000-0002-3886-1258]{Y. Su}
\affiliation{University of Kentucky \\
Lexington, KY, USA}

\begin{abstract}

Future microcalorimeter X-ray observations will resolve spectral features in unmatched detail. 
Understanding the line formation processes in the X-rays deserves much attention. 
The purpose of this paper is to discuss such processes in the presence of a photoionizing source. 
Line formation processes in one and two-electron species are broadly categorized into four cases. 
Case A occurs when the Lyman line optical depths are very small and photoexcitation does not occur. 
Line photons escape the cloud without any scattering. 
Case B occurs when the Lyman-line optical depths are large enough for photons to undergo multiple scatterings.
Case C occurs when a broadband continuum source strikes an optically thin cloud. 
The Lyman lines are enhanced by induced radiative excitation of the atoms/ions by continuum photons, also known as continuum pumping. 
A fourth less-studied scenario, where the Case B spectrum is enhanced by continuum pumping, is called Case D. Here, we establish the mathematical foundation of Cases A, B, C, and D in an irradiated cloud with Cloudy. We also show the total X-ray emission spectrum for all four cases within the energy range 0.1 - 10 keV at the resolving power of XRISM around 6 keV. 
Additionally, we show that a combined effect of electron scattering and partial blockage of continuum pumping reduces the resonance line intensities. Such reduction increases with column density and can serve as an important tool to measure the column density/optical depth of the cloud.

\end{abstract}


\section{Introduction}
Microcalorimeter X-ray missions 
like Hitomi and the upcoming missions \textit{XRISM} and \textit{Athena} will provide unprecedented spectroscopic resolution.
Soft X-ray Spectrometer  \citep[SXS,][]{2016SPIE.9905E..0VK} onboard \textit{Hitomi} \citep{2016Natur.535..117H} resolved the Fe XXV K$\alpha$ complex in four components for the first time. 
A plethora of high-resolution X-ray data from these missions will be available within the next few decades.
Interpreting these high-resolution spectra requires a clear understanding of the line formation processes in the X-ray emitting plasma. 

Line formation in gaseous nebulae was first studied in the 1930s  in a series of papers by
\citet{1937ApJ....85..330M}, \citet{1937ApJ....86...70M}, \citet{1938ApJ....88...52B}, and \citet{1938ApJ....88..422B} for the formation of optical HI lines. Two limiting cases were discussed- ``Case A"  and ``Case B".  
 Case A  occurs when the nebula is optically thin, and the line photons emitted by recombination escape the cloud freely. Case B occurs if the nebula is optically thick, and the line photons scatter multiple times in the cloud. Higher-order Lyman lines are converted into Balmer and Ly$\alpha$ (or K$\alpha$) photons or two-photon continuum. Note that in their study, the source of the radiation was assumed to be of stellar origin. Stars in the gaseous nebulae might have strong Lyman absorption lines in the Spectral Energy Distribution (SED), and there is almost no continuum pumping. This will be relevant to the discussion later.

A third case occurring in optically thin irradiated clouds, ``Case C" , was introduced by  \citet{1938ApJ....88..422B} and later followed up by \citet{1953ApJ...117..399C} and \citet{1999PASP..111.1524F}. In Case C, lines escape the cloud freely like Case A. But unlike Case A, Case C spectrum is enhanced by continuum pumping. 

Some recent studies \citep{2009ApJ...691.1712L, 2016RMxAA..52..419P} discussed a fourth case, `Case D", which occurs in optically thick irradiated systems. Similar to Case B, line photons scatter multiple times in Case D before escaping the optically thick cloud. But unlike Case B, Case D spectrum is enhanced by continuum pumping. 



Most of the previous works on Case A, B, C, and D, both theoretical and observational,  were focused on the optical, ultraviolet, and infrared regime \citep{1937ApJ....85..330M, 1937ApJ....86...70M,1938ApJ....88...52B,1938ApJ....88..422B, 1953ApJ...117..399C, 1981ApJ...243..369S, 1982ApJ...254...22M, 1987MNRAS.224..801H, 1991ApJ...383..135K, 1999PASP..111.1524F, 2007A&A...465..207S, 2012A&A...537A..94S, 2016MNRAS.455.1728M, 2017PASP..129h2001P}, with a small number of studies on the X-rays - \citet{1988MNRAS.231.1139S} and \citet{1995MNRAS.272...41S} for one-electron and   \citet{2007ApJ...664..586P} for two-electron ions. Some other previous studies on soft X-ray spectrum are \citet{2003ARA&A..41..291P}, \citet{2005MNRAS.360..380B},
 \citet{2006A&A...446..459C}, \citet{2007MNRAS.374.1290G}, \citet{2018A&A...612A..18M}. Note that, 
 %
\citet{2002ApJ...575..732K} outlined many of the physical processes discussed in this paper, focusing on second and third-row elements. 

The purpose of our paper is to describe improvements to the widely disseminated code Cloudy with  simultaneous radiative transfer and ionization solutions. This paper presents diagnostic diagrams making it possible to measure column densities from line intensities. Here, we discuss the line formation processes for the four Menzel and Baker cases -- Case A, B, C, and D for H-like and He-like iron in photoionized plasma. This will be essential for interpreting
the future high-resolution microcalorimeter observations in the presence of a photoionizing source.


This paper is the third of the series ``X-ray spectroscopy in the microcalorimeter era", the first two papers of which discussed the atomic processes in a collisionally excited plasma. 
\citet{2020ApJ...901...68C} discussed line interlocking and Resonant Auger Destruction \citep{1978ApJ...219..292R, 1990ApJ...362...90B, 1996MNRAS.278.1082R,2005AIPC..774...99L}, and electron scattering escape (ESE) in the Fe XXV K$\alpha$ complex.
\citet{2020ApJ...901...69C} discussed Case A to B transition in H- and He- like iron. The present paper explores photoionized X-ray plasma with Cloudy \citep{2017RMxAA..53..385F} for a power-law SED. 
Note that, the results shown in all three papers of this series apply in the coronal limit, although the formalism will go to equilibrium in high densities. Figure 10, 11, and 12 in \citet{2017RMxAA..53..385F} display the coronal limit for collisionally ionized and photoionized cases. For iron (Z=26), the coronal limit applies for electron densities smaller than  10$^{16}$ cm$^{-3}$. 


The organization of this paper is as follows. Section \ref{Theory} discusses the theoretical framework of Case A, B, C, and D.
Section \ref{Simulation Parameter} lists the simulation parameters used for our calculations. Section \ref{Results} describes the results.
Section  \ref{Spectral Features}  discusses  the total emitted spectrum within the 
energy range 0.1 - 10 keV.
 Section \ref{effects of continuum opacities} describes the effects of background continuum opacities like electron scattering opacity. Section \ref{summary} discusses our results. We refer to the transitions going from $n = 2, 3, 4$ to $n=1$ in H-like iron as Ly$\alpha$, Ly$\beta$, and  Ly$\gamma$ and in He-like iron as K$\alpha$, K$\beta$, and  K$\gamma$. Transitions going from $n=3$ to $n=2$ are called H$\alpha$ in H-like iron and L$\alpha$ in He-like iron. This nomenclature is inspired by Seigbahn notation \citep{1916Natur..96R.676S} as implemented in, for instance, \citet{1972SSRv...13..655G}, \citet{2004ApJ...613..700F},
 and \citet{10.1093/pasj/59.sp1.S245}.





\begin{figure}[h!]
\centering
\includegraphics[scale=0.5]{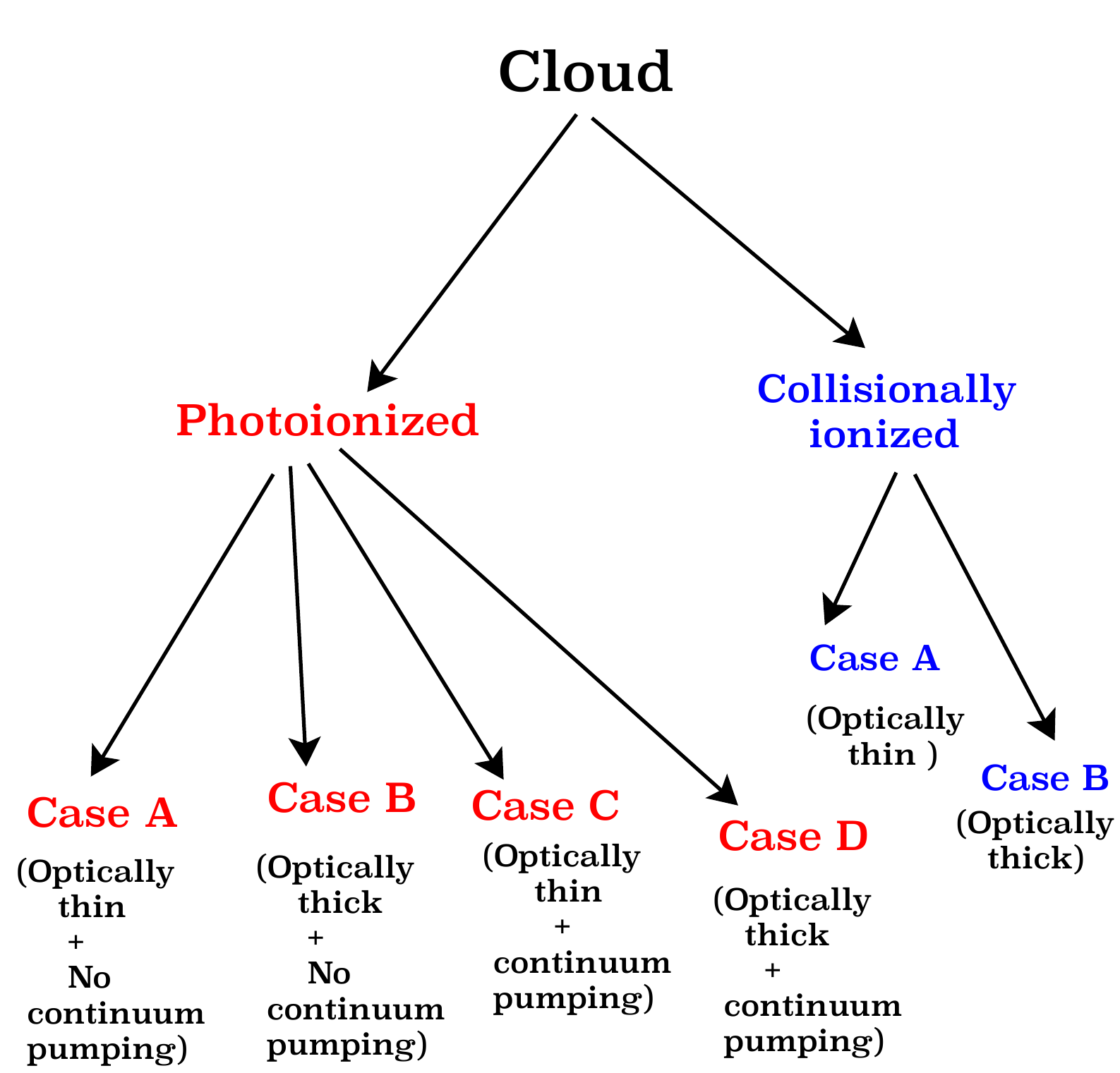}
\caption{A flow-chart showing the line formation conditions occurring in photoionized/collisionally ionized clouds. Case A, Case B, Case C, and Case D occur in photoionized clouds. Case A and Case B also occur in collisionally ionized clouds.
\label{f:flowchart}}
\end{figure}

\begin{figure*}
\gridline{\fig{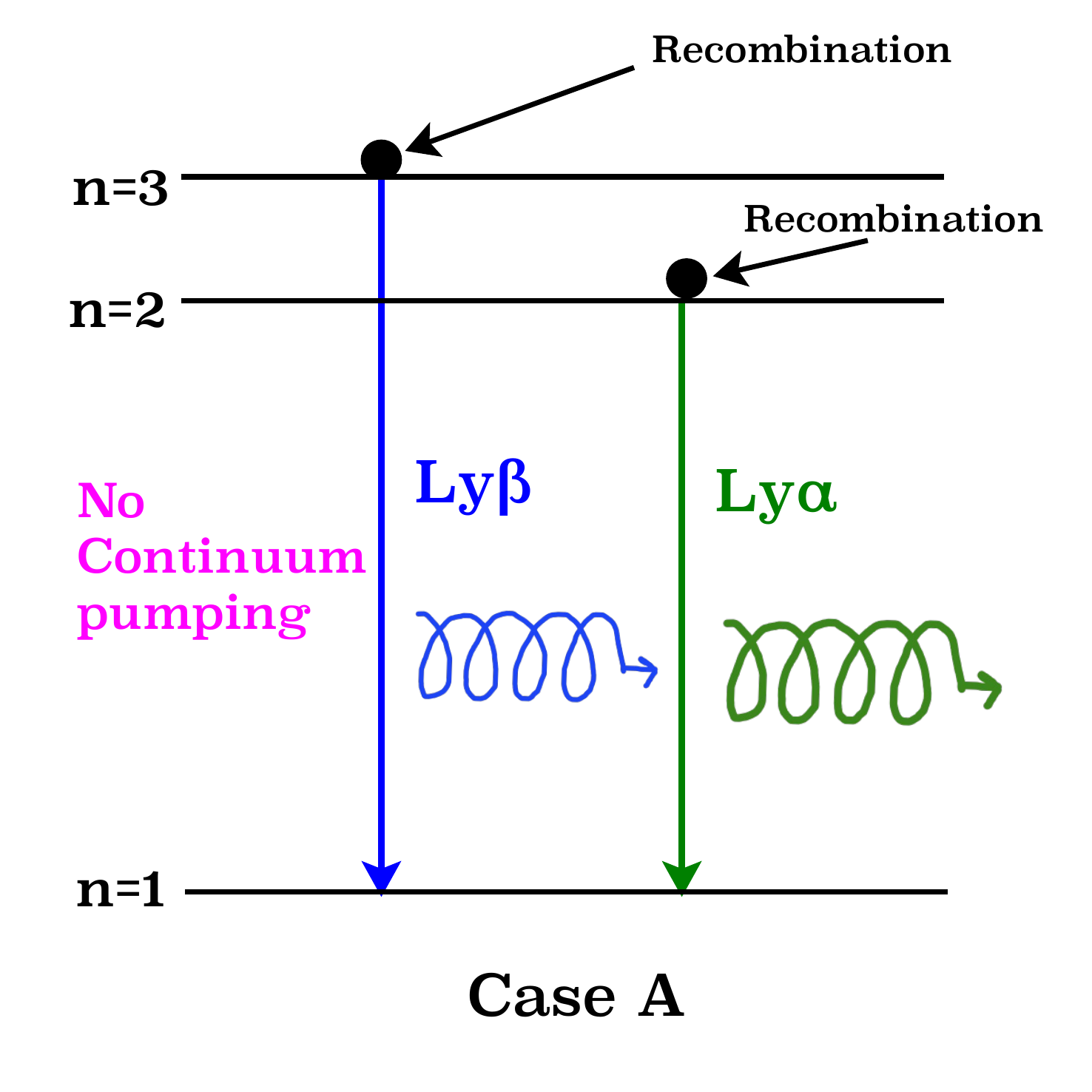}{0.5\textwidth}{(a)}
          \fig{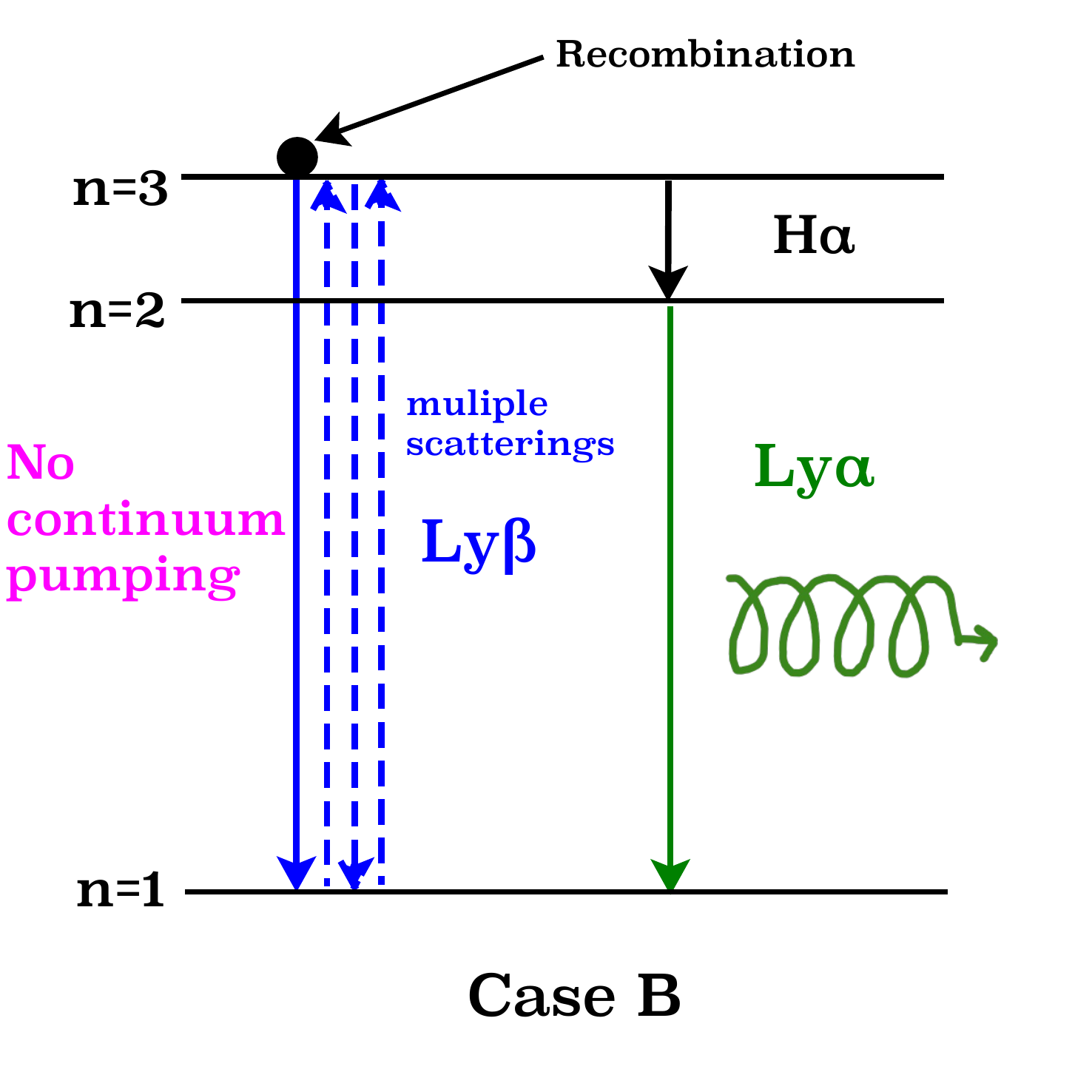}{0.5\textwidth}{(b)}
          }
\gridline{\fig{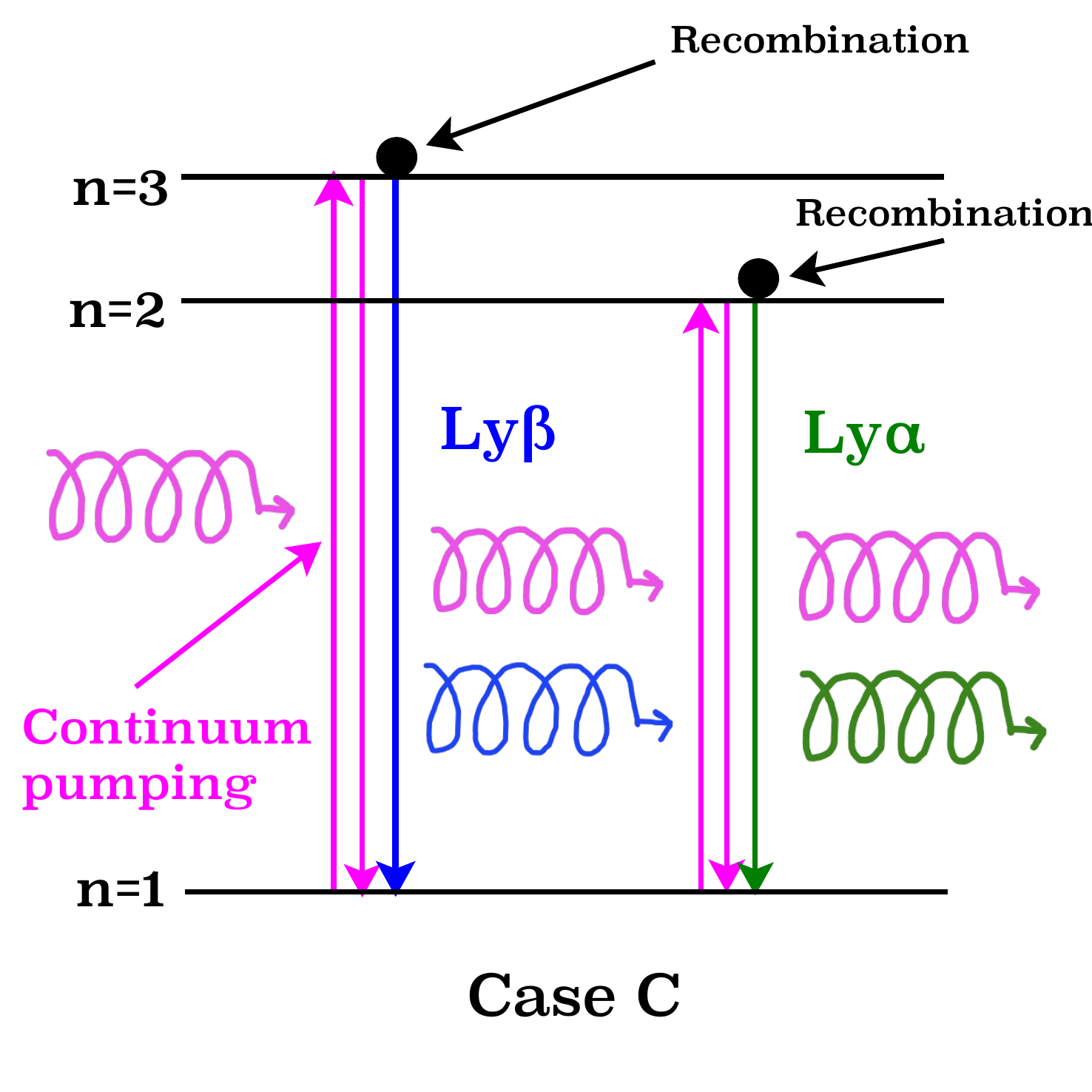}{0.5\textwidth}{(c)}
           \fig{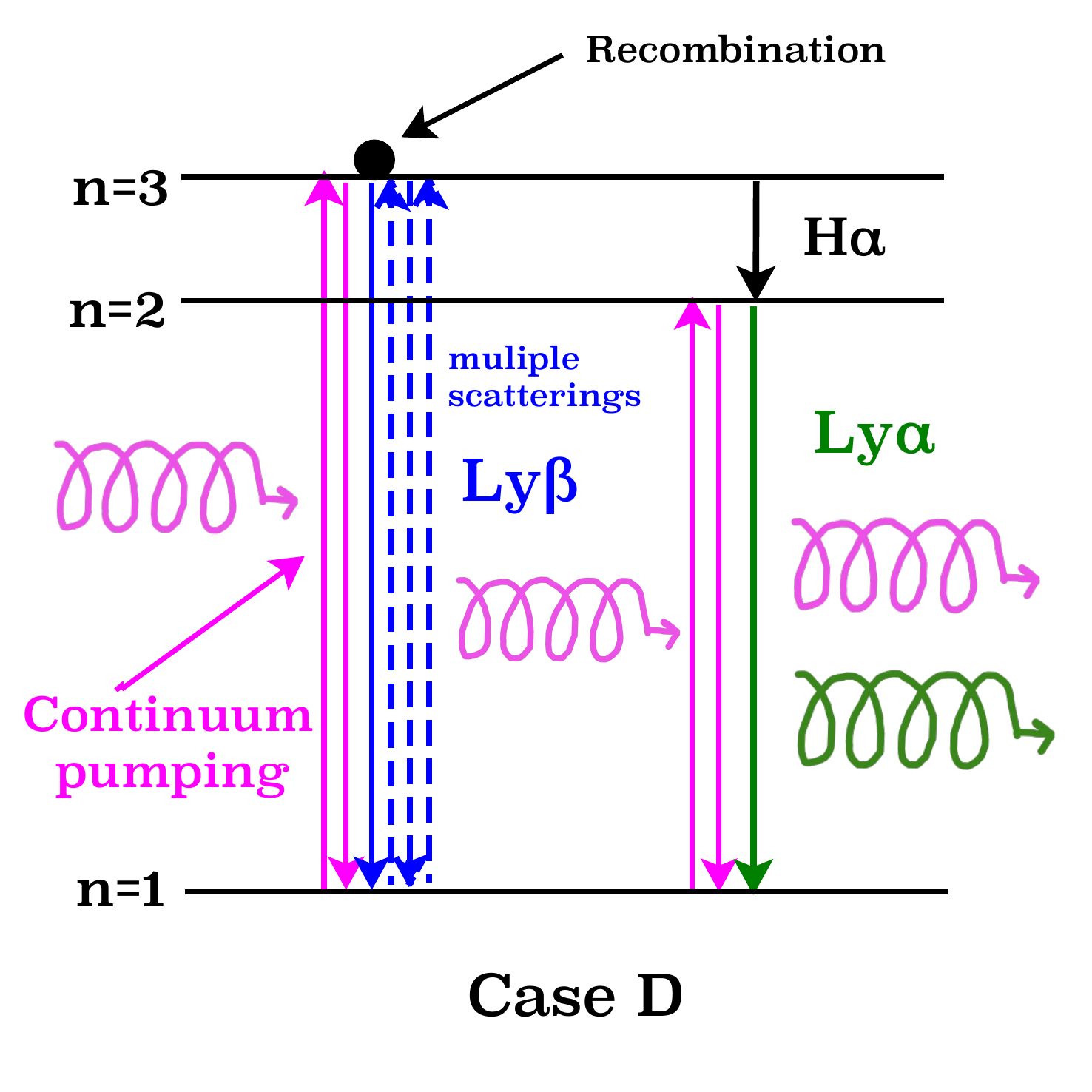}{0.5\textwidth}{(d)}
           }
\caption{ A simplified three-level representation of Case A, Case B, Case C, and Case D shown for a H-like system. The same atomic processes occur in He-like systems too. The top-left panel shows Case A, where the  lines  are  formed by radiative recombination and cascades from higher levels. There is no continuum pumping. The cloud is optically thin, and all Lyman photons escape the cloud without scattering/absorption. The top-right panel shows Case B, where the cloud is optically thick, and continuum pumping photons are blocked. Higher n-Lyman photons ( Ly$\beta$ in the diagram) scatter multiple times to generate Balmer (here H$\alpha$) and Ly$\alpha$ photons.
The bottom-left panel shows Case C, where in addition to radiative recombination and downward cascade, Lyman lines are also enhanced by continuum pumping. The enhanced Lyman lines escape the optically thin cloud. The bottom-right panel shows Case D, where  multiple scatterings and continuum pumping in Lyman lines occur together in an optically thick cloud. Two-photon-continuum is not shown because it does not make emission lines.
\label{f:catoonabc}}
\end{figure*}

\section{Theoretical Framework}\label{Theory}

Lines are formed under Case A, Case B, Case C, or Case D conditions. Case A and Case B occur in collisionally ionized clouds. Case A, Case B, Case C, and Case D occur in photoionized clouds. Figure \ref{f:flowchart} shows the four cases for the two ionizing conditions. Line formation processes in collisonally ionized clouds have been discussed in the first two papers of this series. This paper solely focuses on photoionized clouds. 

A schematic representation of all four cases is shown in Figure \ref{f:catoonabc} for a simplified three-level system. Small-column-density (optically thin) regions can be described by Case A or Case C depending on the ionizing radiation. In both cases, line photons escape the cloud without any scattering.
Case A occurs when the continuum radiation hitting the cloud has strong absorption features in the Lyman lines. There is no continuum pumping in the Lyman lines emitted by the cloud. Lines in the Case A limit are solely formed by radiative recombination and cascades from higher levels\footnote{As described in \citet{2020ApJ...901...69C}, in a collisionally ionized cloud in the absence of photoionizing radiation line formation in optically thin limit is also described by Case A.}.
Case C occurs if the continuum source striking the optically thin cloud does not contain Lyman absorption lines, and the emitted Lyman lines are enhanced by continuum pumping. 
As a result, Case C spectrum is always brighter than Case A. 

The ratio of the Case C to Case A line intensities can be calculated from the ratio of continuum pumping rate to the rate of recombination ($r$). In equilibrium, $r$ is equal to the rate of photoionization ($\Gamma _n$):
\begin{equation}\label{e:gamma}
{\Gamma _n} = \int_{{\nu _o}}^\infty  {\frac{{{4\pi \,J_\nu }}}{{h\nu
}}\;{\alpha _\nu }\;d\nu } \quad [\mathrm{s}^{-1}]
\end{equation}
where $J_\nu$ [erg cm$^{-2}$  s$^{-1}$ sr$^{-1}$ Hz$^{-1}$ ] is the mean intensity per unit frequency, per unit solid angle of the incident radiation, $\alpha _\nu$ is the photoionization cross-section [cm$^{2}$] for the atom/ion by photons of energy $h{\nu}$. 

The rate of continuum pumping of a line from level $l$ to level $u$ is given by:
\begin{equation}\label{e:gamma_l}
{\Gamma_{l}} = {B_{lu}}{J_{l}} 
= {f_{lu}}\frac{{{\pi e{^2}}}}{{{mc}}}(\frac{{{4\pi \,J_{l} }}}{{h\nu
}})
\quad [\mathrm{s}^{-1}]
\end{equation}
  where B$_{lu}$ is the Einstein coefficient, f$_{lu}$ is the oscillator strength, and the other symbols have their usual meanings.

For a power-law (f$_{\nu}$ $\propto$ $ \nu^{-1}$) model in H-like iron for Lyman-$\alpha$ transition in a simple two-level system:
\begin{equation}\label{e:ratio}
\frac{\Gamma_{l}}{r} \sim 16
\end{equation}

This implies that,  Ly$\alpha$ line intensities are  $\sim 16$ times enhanced in Case C compared to Case A . 
The calculated ratio is approximate, as the real calculation will have many pumping lines and many different branching ratios.
This ratio is approximately in agreement with the line intensities listed in Table \ref{t:1} obtained from our Cloudy simulations, which shows that the  Case C Ly$\alpha$ in  H-like iron is $\sim$ 10 times enhanced than Case A.

In contrast, Case B occurs in the high-column-density limit ($N_{\rm H}$ $\geq$  10$^{21.5}$ cm$^{-2}$). In this limit, Lyman line optical depths are
large enough for  photons to undergo multiple scatterings and so are converted into H$\alpha$ (or L$\alpha$)  and Ly$\alpha$ (or K$\alpha$) photons or two-photon-continuum. 
The cloud becomes self-shielding,  stopping the continuum pumping despite the presence of a 
continuum radiation source in the cloud. Typically, Case B describes the line formation in most 
observed optically thick nebulae \citep{2006agna.book.....O}. 

Case D occurs if Lyman line optical depths are large, but the cloud does not entirely become
self-shielding
to the external radiation. \citet{2009ApJ...691.1712L} argued that a real nebula in the 
optically thick limit 
would be better represented by Case D than Case B. Their study reported a significant contribution of continuum pumping on Balmer emissivity in the HII region. As far as we know, Case D had 
not been studied in the X-ray to date.
Our calculations for the X-ray regime in the optically thick limit show substantial enhancement in the
Lyman and Balmer line intensities in H- and He-like iron.
This will be further elaborated in Section \ref{Case D}.

\begin{figure}
\subfigure{\includegraphics[width = 3in]{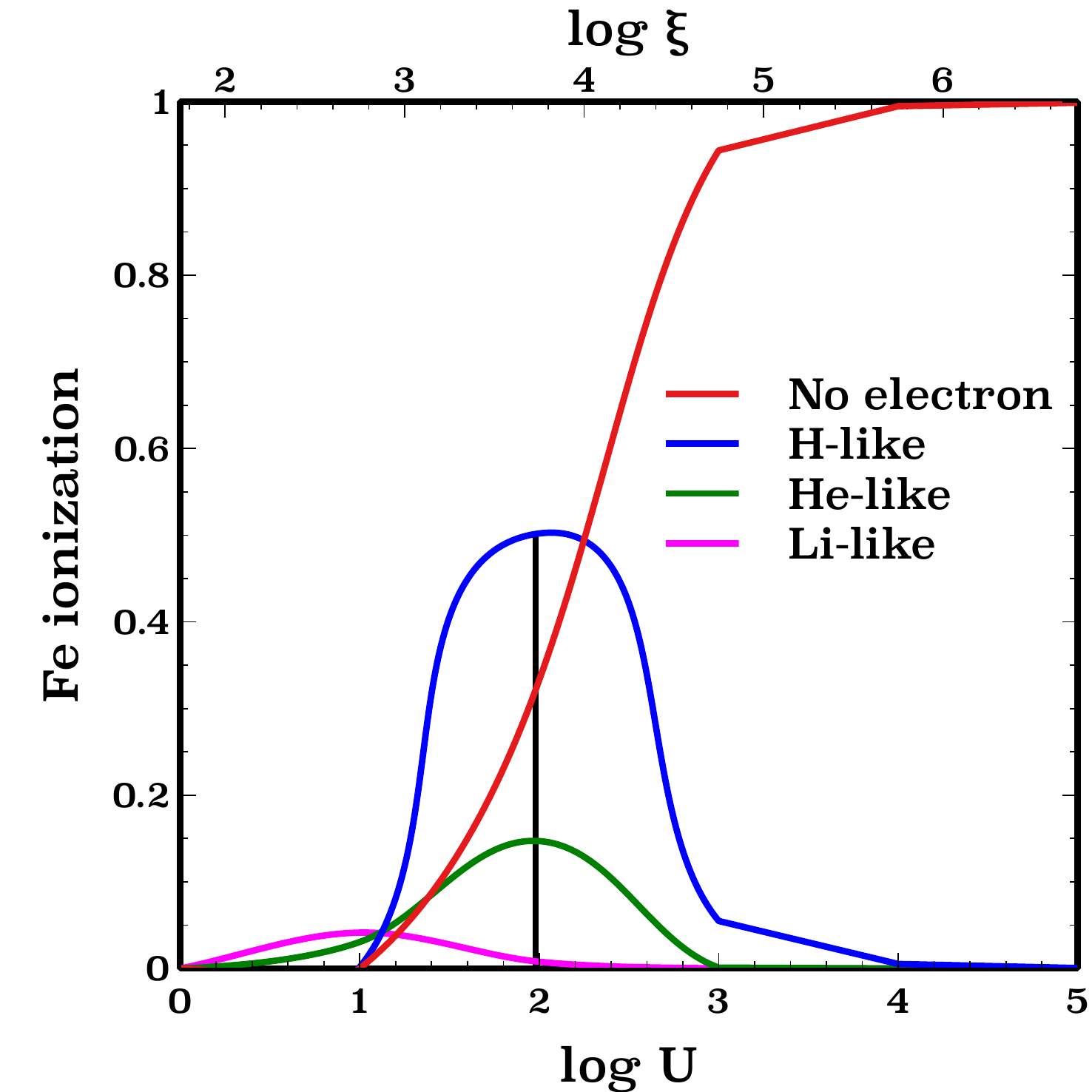}}\hspace{2cm}
\subfigure{\includegraphics[width = 3in]{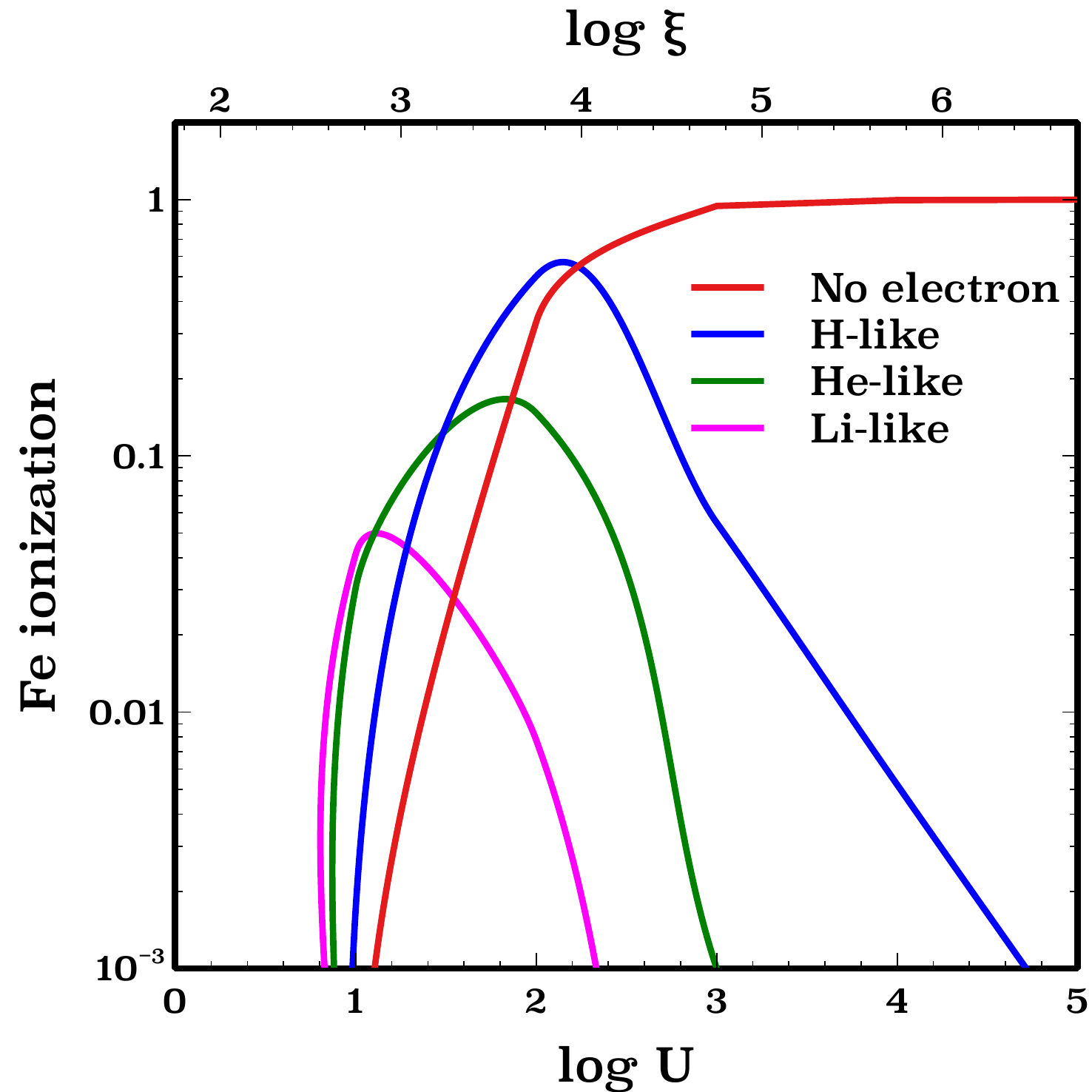}} 
\caption{Ionization in iron versus log of ionization parameter (U). Top and bottom panels show the iron ionization in linear and log scales. Red, blue, green, and magenta lines show the fraction of no-electron, H-like, He-like, and Li-like iron. A vertical black line is drawn at log U = 2 in the linear plot, the ionization parameter chosen for our calculations. The x-axes on the top in both figures show log $\xi$. 
\label{f:1}}
\end{figure}

\begin{figure}[h!]
\centering
\includegraphics[scale=0.5]{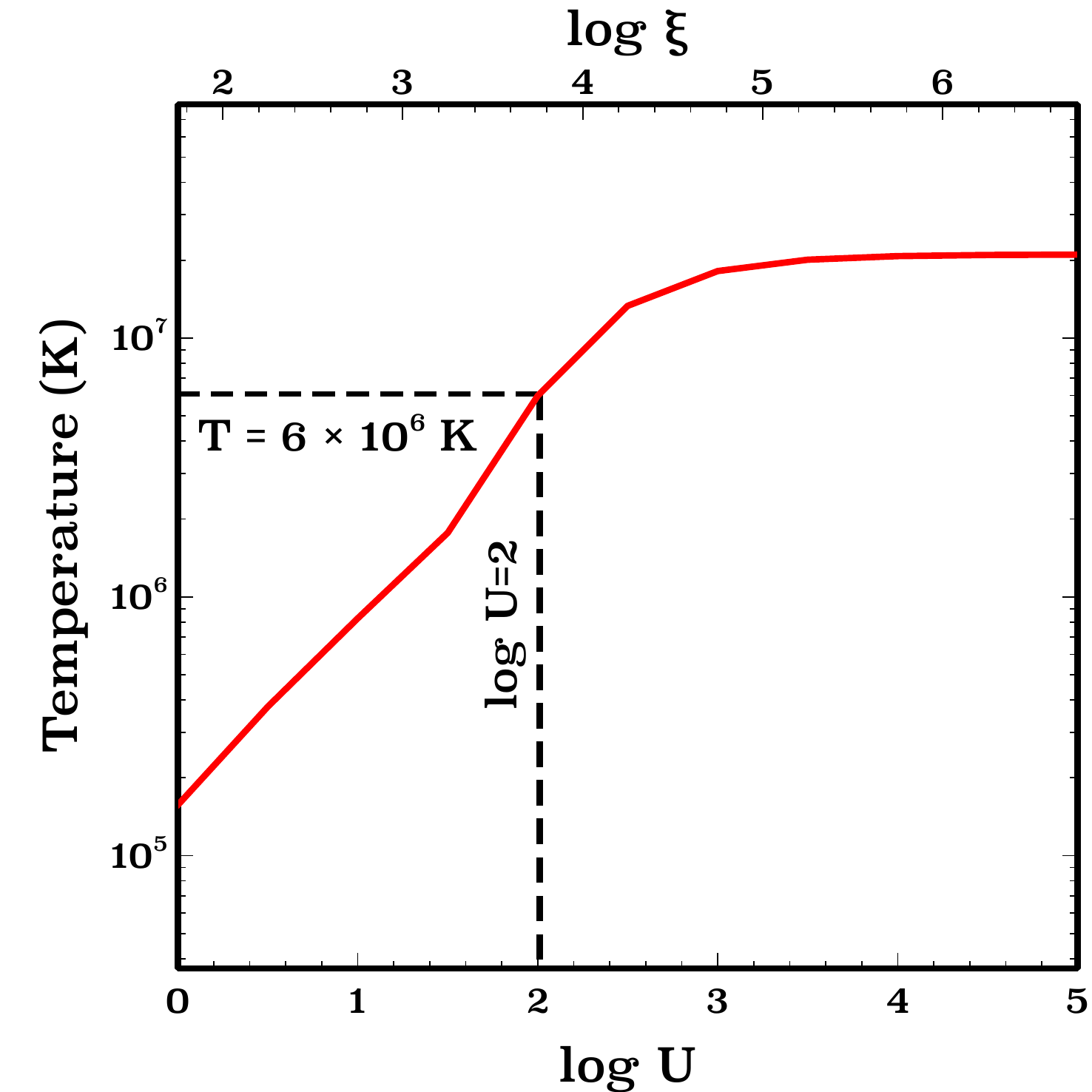}
\caption{Temperature of the photoionized cloud versus log of ionization parameter (log U) for a 1 cm$^{3}$ plasma. The black dashed lines show the temperature at our choice of ionization parameter, log U = 2. The x-axis on the top shows log $\xi$.
\label{f:2}}
\end{figure}
\section{Simulation Parameters}\label{Simulation Parameter}

This section discusses the simulation parameters used in Cloudy. We aim to establish a standard
mathematical framework of line formation through Case A, Case B, Case C, and Case D. The simulation parameters have been  chosen accordingly. All the simulations are done using the
development version of Cloudy with a hydrogen density of 1 cm$^{-3}$. To make the simplest case, 
the shape of the incident radiation field is assumed to be a power-law  spectral energy 
distribution (SED):
\begin{equation}
\rm f_{\nu} \propto \nu^{-\alpha}
\end{equation}
with $\alpha =1$.

The intensity of the radiation field is characterized by ionization parameter (U), defined with the following ratio:
\begin{equation}
U = \frac{\Phi_{H}}{n_{H}\, c}
\end{equation}
where $\Phi_{H}$ is flux of hydrogen-ionizing photons, $n_{H}$ is the hydrogen-density, and c is the speed of light. 

Much of the X-ray literature uses the $\xi$ ionization parameter defined by \citet{2001ApJS..133..221K}.  For our $\alpha$ =1 SED,  $\xi$ = 1 corresponds to an U of 0.01767.

Figure \ref{f:1} shows the variation of different ionization stages in iron with the log of U. The top x-axis shows the log of $\xi$.  The top panel of the figure shows a linear plot, and the bottom panel shows a log plot. We choose log U=2 (log $\xi$ = 3.75) to maximize the quantity of
H- and He-like iron in the cloud. This also minimizes the overlap between He- and Li-like iron ions, as Li-like iron selectively changes the He-like spectra by line interlocking \citep{2020ApJ...901...68C}. 
The linear figure also shows our choice of ionization parameter with a black vertical line.
Note that, our choice of ionization parameter is in agreement with the range of $\xi$  mentioned  in \citet{2004ApJS..155..675K} (log $\xi$ $\geq$ 2) for highly charged iron, who also discussed K lines in iron in a photoionized cloud for a power-law SED.

The cloud temperature (T) is obtained from the energy equilibrium equation from a heating cooling balance to replicate the actual physical temperature of a photoionized cloud.  The computed temperature is - T $\sim$ 6 $\times$ 10$^{6}$ K at log U = 2.
The variation of T with log U is shown in Figure \ref{f:2}. Note that, photoionized clouds are a lot cooler than collisionally ionized clouds. In fact, this equilibrium temperature is $\sim$ 8 times smaller than that of the collisionally ionized plasma considered in the first two papers of this series.

\begin{figure*}
\gridline{\fig{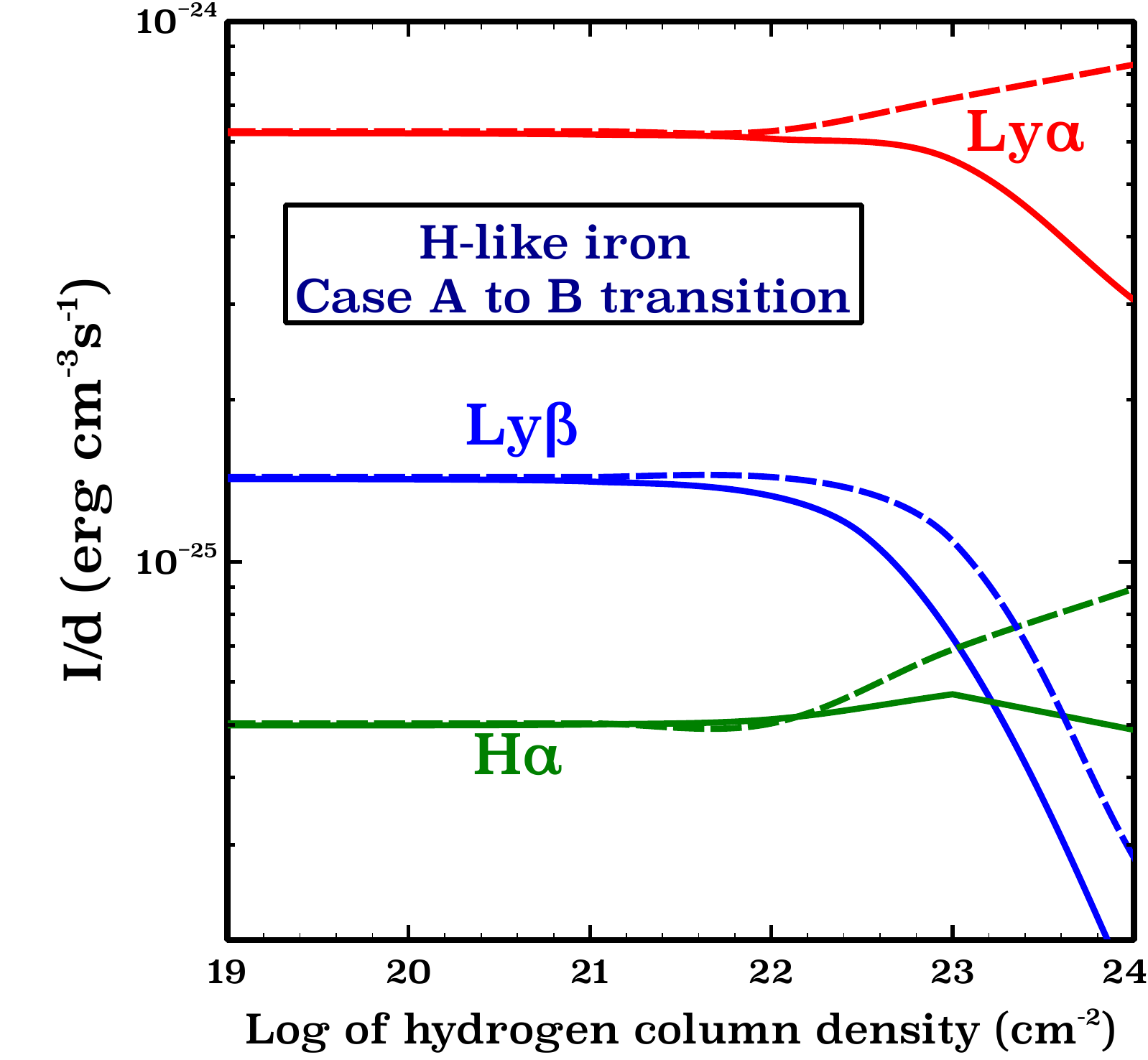}{0.45\textwidth}{(a)}
          \fig{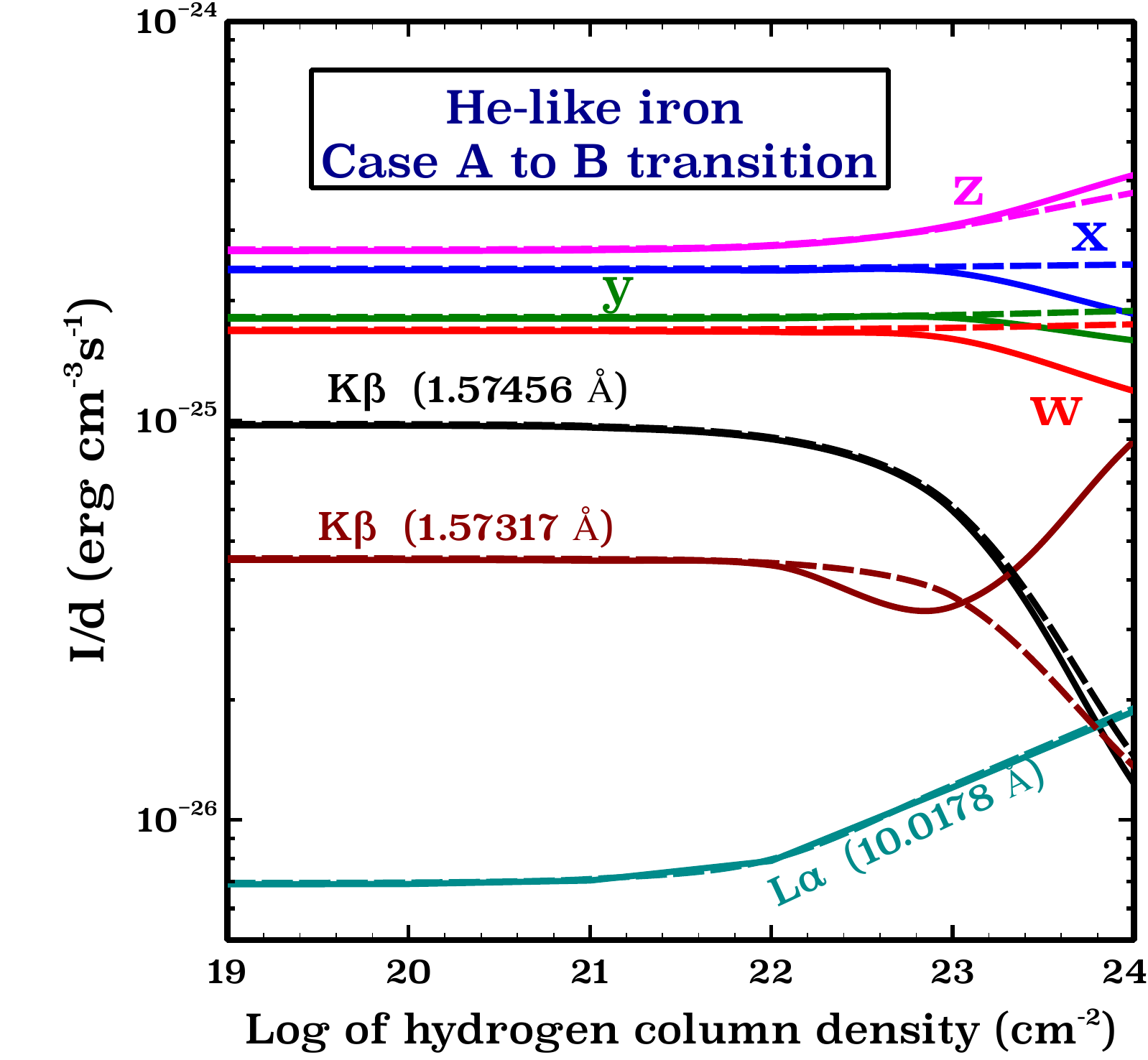}{0.45\textwidth}{(b)}
          }
\gridline{\fig{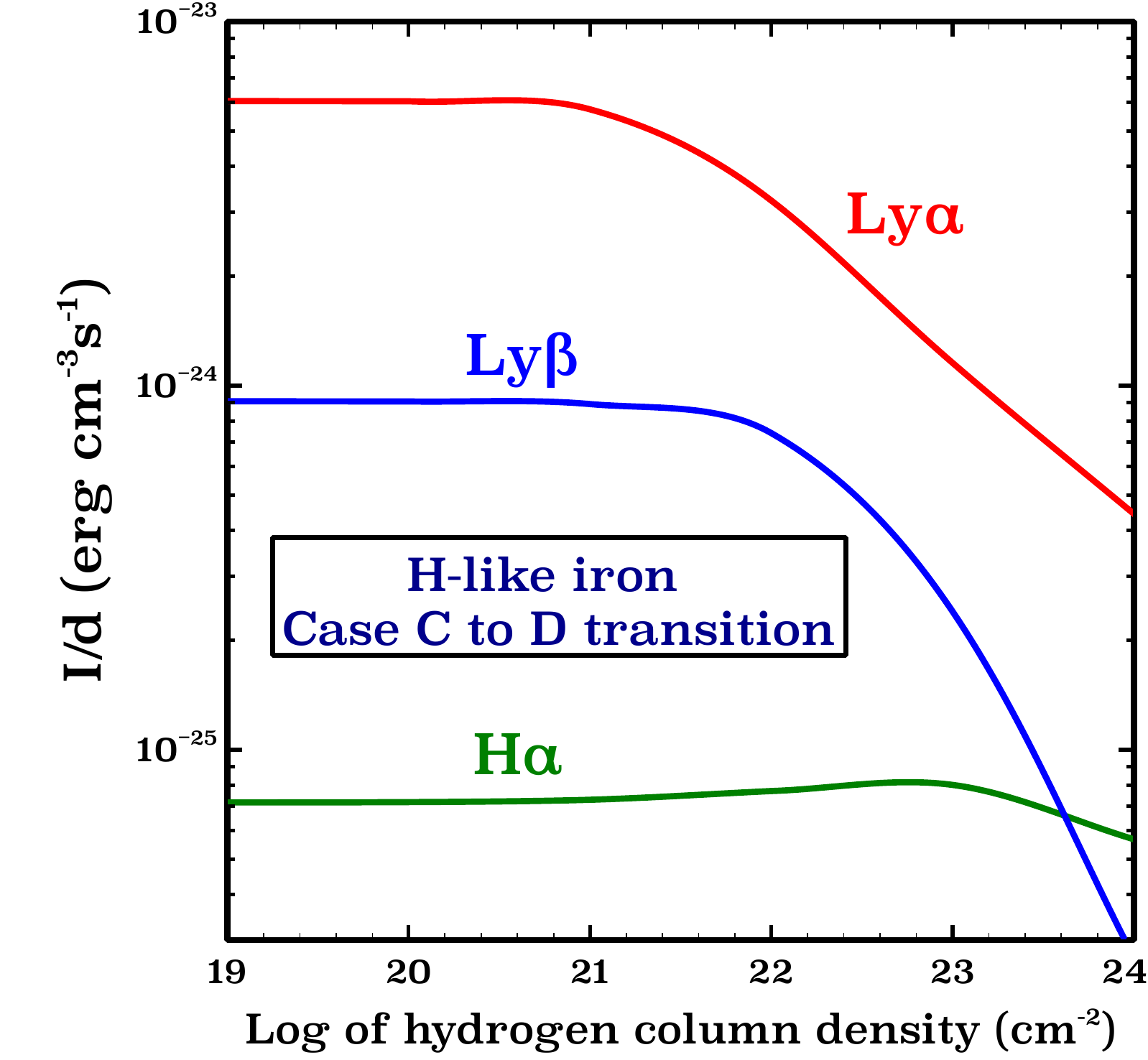}{0.45\textwidth}{(c)}
          \fig{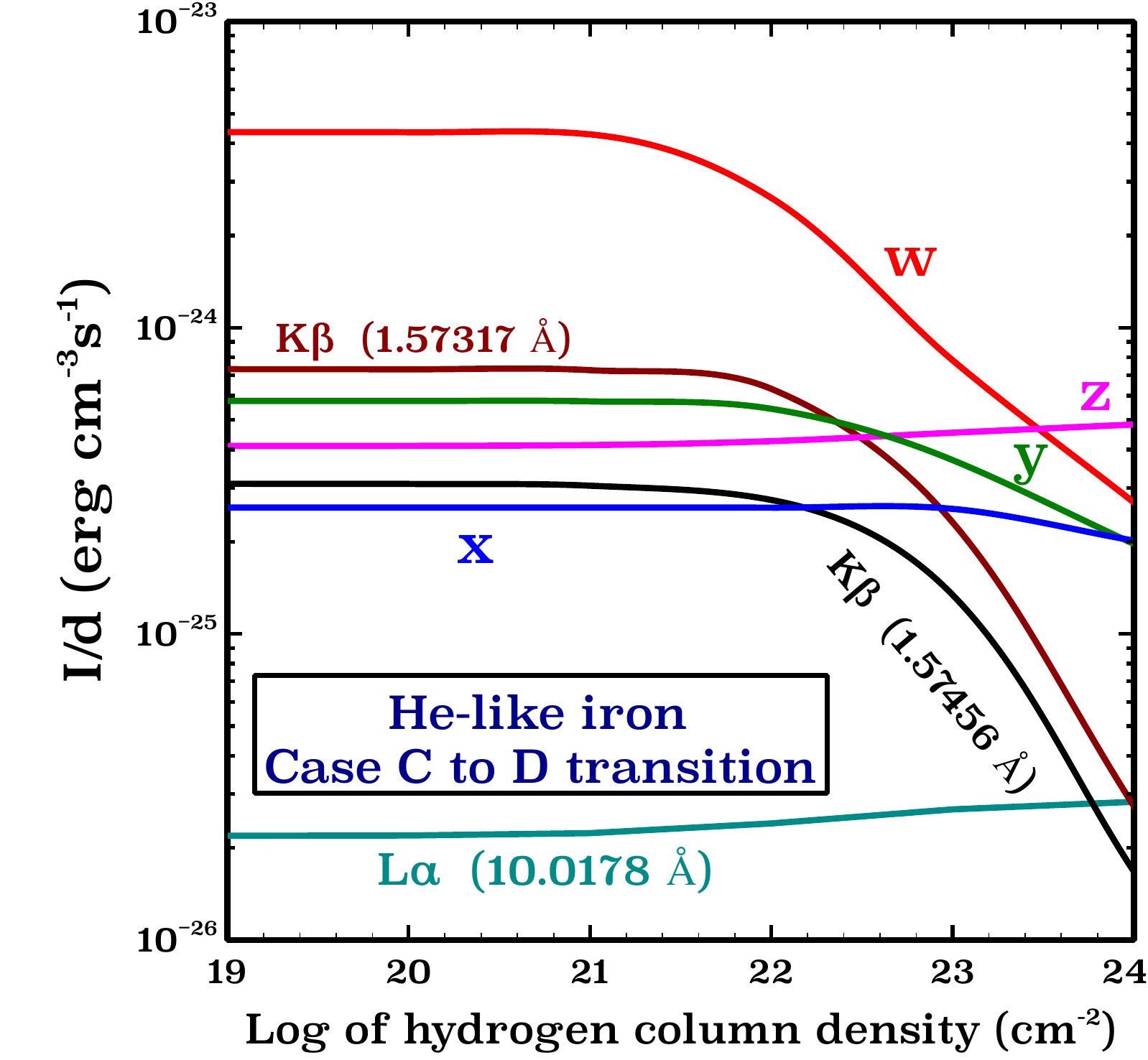}{0.45\textwidth}{(d)}
          }
\caption{Line intensity per unit thickness of the cloud versus the log of hydrogen column density. Figures in the top panel show Case A to Case B transition in H- and He-like iron. Dashed lines represent the classic Menzel-Baker Case A to Case B limit, solid lines represent the observed Case A to Case B limit.  Figures in the bottom panel show the observed Case C to Case D transition in H- and He-like iron. 
\label{f:col_den}}
\end{figure*}

\begin{table*}[ht]
\centering
\caption{Line intensities per unit thickness (I/d) for Case A, B, C, and D conditions for certain Lyman and Balmer transitions in H- and He-like ions in a photoionized cloud. I/d's for Case A and C are listed for N$_{H}$=10$^{19}$ cm$^{-2}$ and for Case B and D are listed for N$_{H}$ = 10$^{24}$ cm$^{-2}$. The I/d's listed under Case A, Case B, Case C, and Case D are what will be observed in nature. The Case B$_{\rm classic}$, the classic Menzel-Baker Case B, has also been included in the table for educational purposes.}
\begin{tabular}{lllrrrrr}
\hline
\multicolumn{6}{r}{I/d [erg cm$^{-3}$ s$^{-1}$]} \\
\cline{4 -8}
Ion & Transitions &Wavelength   & Case A & Case B$_{\rm classic}$ & Case B &  Case C & Case D \\
\hline
 & $2^{2}P\rightarrow1^{2}S$& 1.77982\AA & 6.22e-25  &  8.24e-25  & 3.05e-25  & 6.05e-24  & 4.15e-25  \\
&$3^{2}P\rightarrow1^{2}S$ & 1.50273 \AA &1.43e-25 &  2.84e-26 &  1.57e-26 & 9.06-25 & 2.69e-26 \\
H-like& $3^{2}P\rightarrow2^{2}S$ & 9.65247 \AA & 4.99e-26 &  8.94e-26 &  4.89e-26 & 7.17e-26 &  6.68e-26 \\
& $4^{2}P\rightarrow1^{2}S$& 1.42505 \AA &   5.55e-26 & 1.05e-26   &  9.67e-27  &  3.07e-25 & 1.89e-26  \\
& $4^{2}P\rightarrow2^{2}S$& 7.14920 \AA &   1.86e-26 & 3.09e-26   &  1.69e-26  &  2.71e-26 & 2.05e-26  \\
\hline
 &$2^{1}P\rightarrow1^{1}S$(w)&1.85040 \AA &1.69e-25 & 1.74e-25 &  1.29e-25 &  4.35e-24 & 2.69e-25 \\
 &$2^{3}P_{2}\rightarrow1^{1}S$(x)&1.85541 \AA & 2.41e-25 &  2.46e-25 &  1.85e-25 &  2.59e-25  & 2.02e-25 \\
 & $2^{3}P_{1}\rightarrow1^{1}S$(y)&1.85951 \AA & 1.82e-25 & 1.89e-25  & 1.59e-25 &  5.77e-25 &  1.97e-25 \\
He-like& $2^{3}S\rightarrow1^{1}S$(z) &1.86819 \AA & 2.68e-25 &  3.72e-25 &   4.23e-25 &  4.12e-25 &  4.83e-25 \\
& $3^{1}P\rightarrow1^{1}S$ &1.57317 \AA &  4.53e-26 &  1.36e-26 & 8.89e-27 &  7.32e-25 & 2.74e-26\\
& $3^{3}P\rightarrow1^{1}S$ &1.57456 \AA & 9.81e-26  &  1.45e-26 & 1.23e-26 & 3.09e-25 & 1.68e-26 \\
& $3^{1}P\rightarrow2^{1}S$ &10.2202 \AA & 4.41e-28 & 5.53e-27 & 5.41e-27 &  7.14e-27 &  9.72e-27\\
& $3^{3}P\rightarrow2^{3}S$ &10.0178 \AA &  6.96e-27 & 1.91e-26 &  1.87e-26 & 2.19e-26 & 2.82e-26 \\
& $4^{1}P\rightarrow1^{1}S$ &1.49460 \AA &  1.82e-26 & 8.28e-27 &  5.37e-27 & 2.51e-25 & 1.92e-26\\
& $4^{3}P\rightarrow1^{1}S$  &1.49513 \AA & 3.61e-26 &  9.36e-27 & 7.77e-27 & 1.06e-25 & 1.19e-26\\
& $4^{1}P\rightarrow2^{1}S$ &7.61825 \AA &  2.40e-28 & 1.73e-27 & 1.68e-27 & 3.32e-27 &  3.77e-27\\
& $4^{3}P\rightarrow2^{3}S$ &7.48713 \AA &  3.24e-27 & 6.91e-27 & 6.78e-27 & 9.54e-27 & 1.11e-26\\
\hline
\end{tabular}
\label{t:1}
\end{table*}

\section{Results}\label{Results}
This section explores different conditions for the lines to form in the presence of a photoionizing source emitting in X-rays. Table \ref{t:1} shows a comparison between selective line intensities (including the K$\alpha$ complex) for Case A, Case B, Case C, and Case D conditions for H- and He-like iron. The quantity listed in the table is the line intensity (I) per unit thickness (d), I/d. As the line intensities increase as the cloud's size/thickness increases, I/d tracks the scaled change in the line intensities for all four cases or the transition between them. Although I/d has the same units as the emissivity of a line, 4$\pi j$, it does not have the same physical interpretation. All the I/d's listed in Table \ref{t:1} are observed in nature except for the Case B$_{\rm classic}$, which will be further discussed in section \ref{Case B}. Line wavelengths listed in the table are taken from  NIST\footnote{\url{ https://physics.nist.gov/asd } } \citep[version 5.8:][]{2018APS..DMPM01004K}.

The low-column-density (optically thin) limit represents Case A and C, and the high-column-density (optically thick) limit represents Case B and D. Therefore, I/d's listed in the table for Case A and C are computed at N$_{H}$=10$^{19}$ cm$^{-2}$ and for Case B and D at N$_{H}$=10$^{24}$ cm$^{-2}$.
The continuous variation of I/d with hydrogen column density and the transition from Case A to B and Case C to D are shown in Figure \ref{f:col_den}.

\subsection{Case A}\label{Case A}
The schematic representation of Case A is shown in the upper-left panel of Figure \ref{f:catoonabc}. Case A is the simplest of all four cases occurring in optically thin systems. As mentioned in the introduction, Case A was developed for SEDs with strong Lyman absorption features, for example, stellar SEDs. The strong Lyman absorption lines in the SED prevent the continuum pumping.
The continuum pumping in our simulated cloud is stopped using the cloudy command:
\begin{verbatim}
no induced processes
\end{verbatim}
 The top panel of Figure \ref{f:col_den} shows that I/d in H- and He-like iron remains constant up to $\sim$ N$_{H}$=10$^{21.5}$ cm$^{-2}$. This is because up to this column density, Lyman lines escape without any scattering/absorption, and any column density smaller than N$_{H}$=10$^{21.5}$ cm$^{-2}$ will generate a pure Case A spectrum. Table \ref{t:1} shows I/d for Case A at N$_{H}$=10$^{19}$ cm$^{-2}$.



\begin{figure}[h!]
\centering
\subfigure{\includegraphics[width = 2.5in]{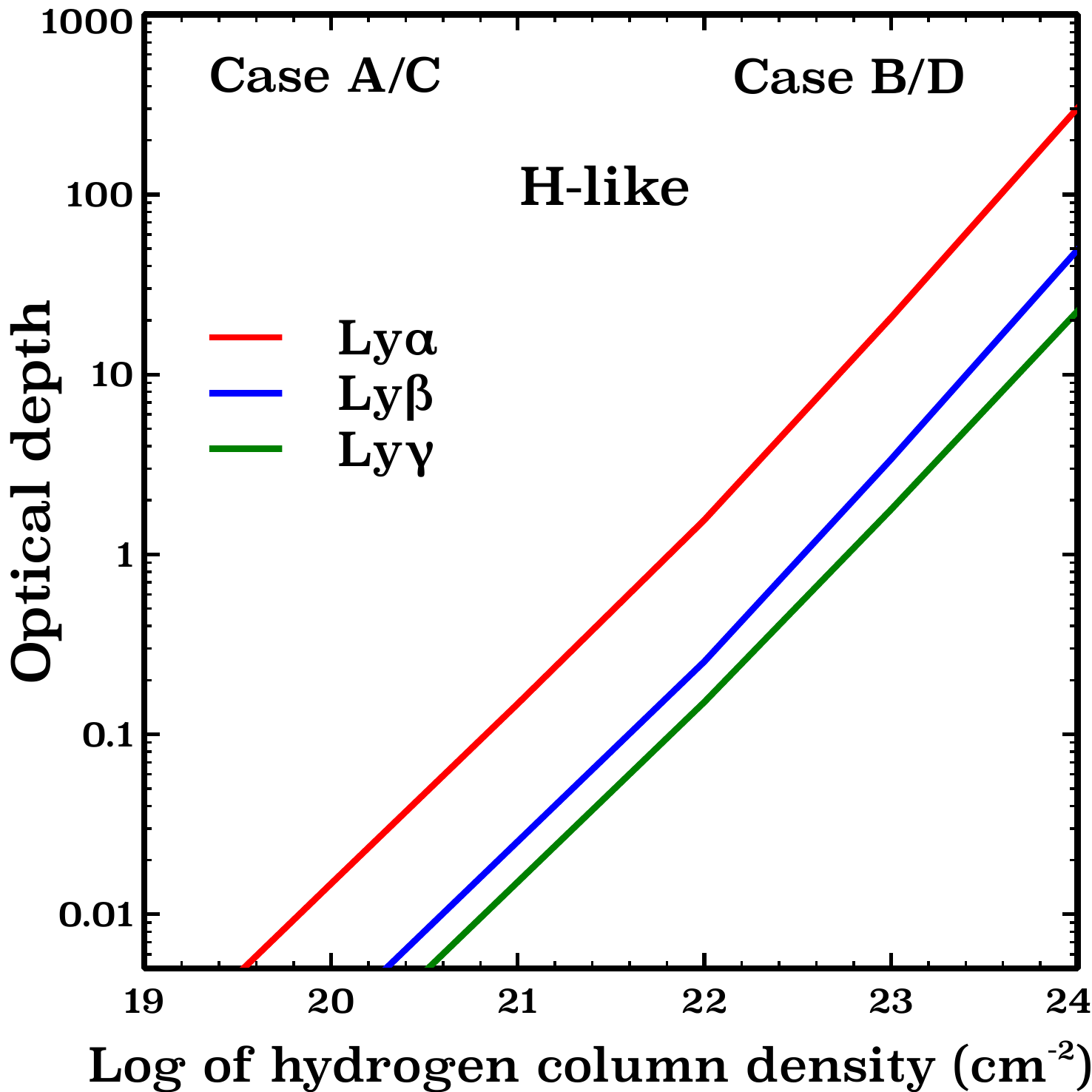}}\hspace{2cm}
\subfigure{\includegraphics[width = 2.5in]{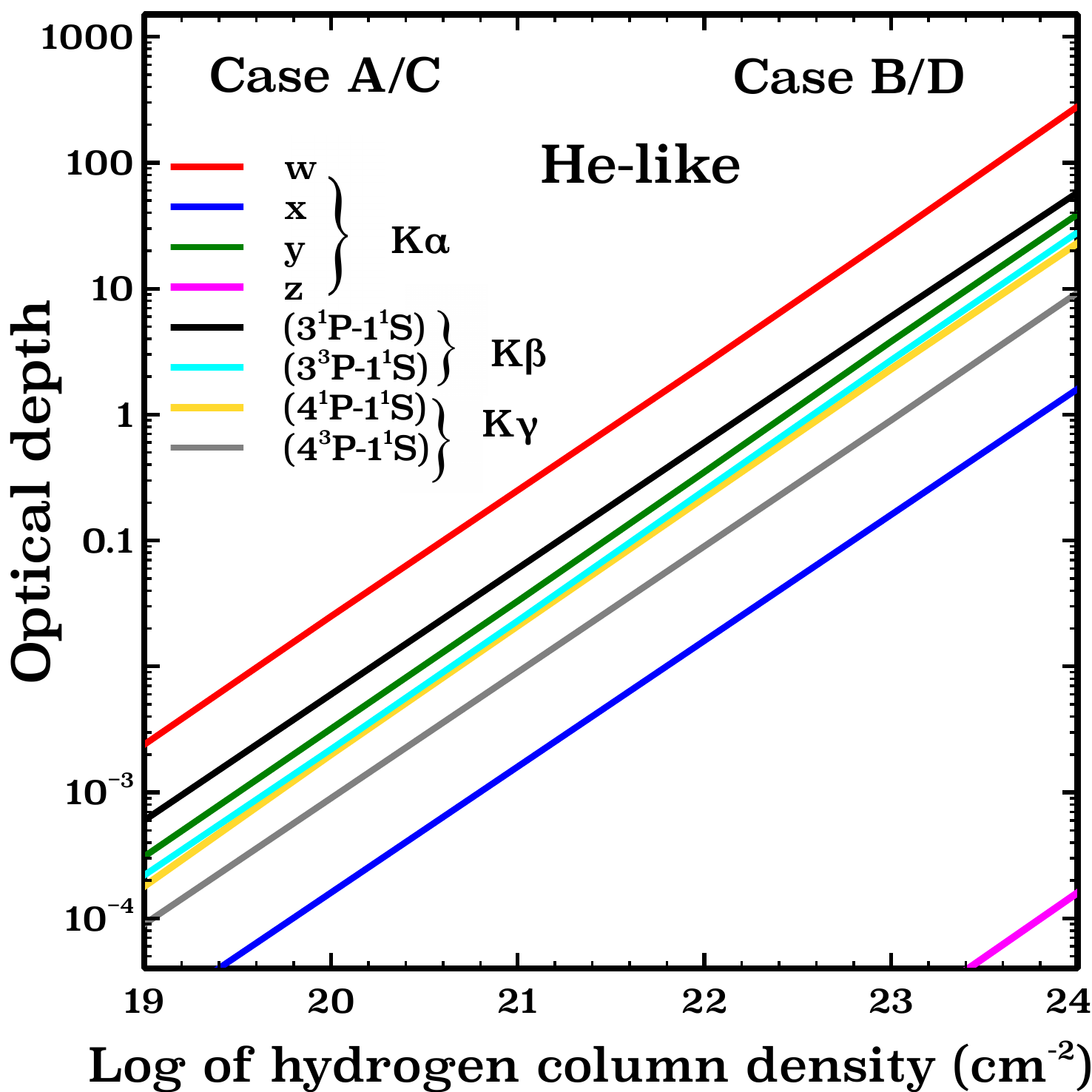}} 
\caption{Optical depth versus log of hydrogen column density for important 
Ly$\alpha$, Ly$\beta$, and Ly$\gamma$ transitions in H- iron, and K$\alpha$, K$\beta$, and K$\gamma$ transitions
in He-like iron. Low-column-density limit represents Case A or Case C. High-column-density limit represents
Case B or Case D.
\label{f:opt}}
\end{figure}

\subsection{Case B}\label{Case B}

The top-right panel of Figure \ref{f:catoonabc} shows the Case B condition in a cloud.
The continuum pumping is disabled as section \ref{Case A}.
As shown in Figure \ref{f:col_den}, the column densities bigger than  N$_{H}$ $\sim$ 10$^{21.5}$ cm$^{-2}$ begin to
make a transition to the Case B limit. Two types of Case B have been shown - Case B$_{\rm classic}$ and Case B. The dashed and solid lines in the top panel of Figure
\ref{f:col_den} represent Case B$_{\rm classic}$ and Case B, respectively. What we refer to as Case B throughout the text considers all the physical processes observed in nature in the X-ray limit, including electron scattering and line overlap. The concept of electron scattering is described later 
in this section and in section \ref{effects of continuum opacities}. 
Case B$_{\rm classic}$ is the Menzel-Baker Case B values studied in the 1930's that does not include electron scattering and line overlap. In Cloudy, electron scattering can be disabled with the following command:
\begin{verbatim}
no electron scattering
\end{verbatim}
We find that, for X-ray emission from H- and He-like iron, Case B$_{\rm classic}$ remains a pedagogical scenario.

Table \ref{t:1} shows the I/d values for Case B$_{\rm classic}$ and Case B (observed in nature) at  N$_{H}$=10$^{24}$ cm$^{-2}$. In the typical Menzel-Baker Case B$_{\rm classic}$, as a consequence of conversion of higher-n Lyman lines into Ly$\alpha$ (or K$\alpha$) and Balmer lines, Ly$\alpha$ (or K$\alpha$) intensity increases, Ly$\beta$ (or K$\beta$) and higher-n Lyman line intensities decrease, and H$\alpha$ (or L$\alpha$) and higher-order Balmer line intensities increase compared to the Case A limit. Case B$_{\rm classic}$ column in the table exactly reflects this behaviour both for H- and He-like iron. For instance in H-like iron, I/d in Ly$\alpha$ increases by $\sim$ 32\%, Ly$\beta$  decreases by $\sim$ 80\%, and H$\alpha$ increases by $\sim$ 80\% at N$_{H}$=10$^{24}$ cm$^{-2}$. In He-like iron, I/d in z increases by $\sim$ 40\%, x , y, and w increase very slightly ($\leq$ 4\%), K$\beta$  decreases by $\sim$ 3 - 7 times, and L$\alpha$ increases by $\sim$ 3 - 13 times compared to Case A.

However, the observed Case B values are quite different from the Case B$_{\rm classic}$ values. In fact, in H-like iron, all Lyman I/d values decrease compared to the Case A limit, including Ly$\alpha$. H-like  Ly$\alpha$ decreases by $\sim$ 50\%.  In He-like iron, selected K$\alpha$ lines (x, y, and w) show a decrease up to $\sim$ 24\%.

Such decrease in the line intensities in the observed Case B mainly occurs due to electron scattering, as described in section \ref{effects of continuum opacities}. When line photons scatter off high-speed electrons, they are largely Doppler-shifted from their line-center. Lines with the largest optical depths are more likely to exhibit a reduction in their line intensities as they are more likely to scatter. Figure \ref{f:opt} shows the optical depth of certain Lyman and Balmer lines in H- and He-like iron. In H-like iron,  Ly$\alpha$,  Ly$\beta$ , and  Ly$\gamma$ intensities are reduced due to electron scattering because of their large optical depths. In He-like iron, w, y,  K$\beta$ , and  K$\gamma$ intensities are reduced. The optical depth in z is negligible at N$_{H}$=10$^{24}$ cm$^{-2}$, and thus I/d for z is not affected by electron scattering. The reduction in I/d for x is not due to electron scattering
but a series of processes described in the appendix of \citet{2020ApJ...901...68C}.
This explains the observed Case B behavior both in Table \ref{t:1} and Figure \ref{f:col_den}.

\subsection{Case C}\label{Case C}
The bottom-left panel of Figure \ref{f:catoonabc} represents the Case C condition in a photoionized cloud. Case C occurs in optically thin clouds, like Case A. However, the Case C spectrum is enhanced by the continuum pumping/fluorescence and makes the brightest spectrum of all four cases. This can be seen in Table \ref{t:1} and the bottom panel of Figure \ref{f:col_den}. The degree of enhancement depends on the shape of the incident radiation field \citep{1999PASP..111.1524F}. In our case, a power-law SED fluoresces the cloud (refer to section \ref{Simulation Parameter} for details). \\
The enhancement in the I/d values in Case C is measured with respect to the Case A spectrum. 
From the Cloudy simulation listed in Table \ref{t:1}, we find that the I/d for Case C  for the Ly$\alpha$ transition for H-like iron gets $\sim$ 10 times amplified compared to that of Case A. This approximately agrees with the theoretical value of amplification shown in section \ref{Theory}. Both Ly$\beta$ and Ly$\gamma$ lines get  amplified by  $\sim$ 6 times.  Further, our calculation for He-like iron shows that w is enhanced $\sim$ 27 times, x, y, and z are enhanced  $\sim$ 1.1, 3.2, and 1.5 times, respectively. The K$\beta$ and K$\gamma$ transitions are enhanced by $\sim$ 3 - 16 times and $\sim$ 3 - 14 times, respectively. The Balmer lines are enhanced up to $\sim$ 16 times.

\subsection{Case D}\label{Case D}
The Case D condition in the cloud is shown in the bottom-right panel of Figure \ref{f:1}, where continuum pumping and multiple scatterings happen together. The cloud is partially self-shielded, and lines are partially enhanced by incident radiation. Case D has been 
hardly discussed in the literature, as at the very high column densities,
a cloud can become entirely self-shielded, and there is essentially no Case D.
The spectral behavior is described by Case B in such cases. 

Case D becomes useful when the cloud's column density is high enough to allow multiple scatterings but can not entirely stop the continuum radiation from penetrating the cloud. 
In fact we find that, for X-ray emission from H- and He-like plasma, Case D deviates considerably from Case B even at a column density as high as N$_{H}$=10$^{24}$ cm$^{-2}$.
As shown in Table \ref{t:1} at  N$_{H}$=10$^{24}$ cm$^{-2}$, the observed Case D value of I/d for H-like iron is $\sim$ 36\% enhanced in Ly$\alpha$,  $\sim$ 71\%  enhanced in Ly$\beta$, and $\sim$ 37\% enhanced in H$\alpha$ compared to the observed Case B value. In He-like iron, w, x, y, and z are enhanced $\sim$ 109\%, $\sim$ 9\%, $\sim$ 24\%, and $\sim$ 14\%, respectively. K$\beta$ is enhanced by $\sim$ 208\% (for $3^{1}P\rightarrow1^{1}S$), and $\sim$ 37\% (for $3^{3}P\rightarrow1^{1}S$), respectively. L$\alpha$ is enhanced $\sim$ 80\% (for $3^{1}P\rightarrow2^{1}S$), and $\sim$ 51\% ( for $3^{3}P\rightarrow2^{3}S$), respectively.

The future microcalorimeters will detect ever-so-slight changes in the spectra, thanks to their unmatched spectral resolution.
Thus it becomes crucial to understand the Case D behavior in optically thick irradiated clouds and its deviation from Case B, especially for column densities  N$_{H}$ $\leq$ 10$^{24}$ cm$^{-2}$. Needless to say,
at even bigger column densities when the optical depth becomes very large the external radiation will be completely absorbed
in the gas. Case D values will eventually approach the Case B values. But until the cloud is thick enough to stop continuum pumping entirely, Case D will be the best description of the emission spectra in irradiated clouds.

\subsection{Case A/Case C to Case B/Case D transition}

What drives the Case A to Case B or Case C to Case D transition  in a real astronomical scenario is the variation in column density from low-column-density (optically thin) to high-column-density (optically thick) limit. Figure \ref{f:col_den} shows these transitions for H-like and He-like iron in a photoionized cloud.

When Case A and B were first discussed in the 1930s, 
the source of the radiation was assumed to be stellar with strong Lyman absorption lines and no continuum pumping. Thus, galactic nebulae with strong absorption features show Case A to Case B transition under the variation in column density. Case A to Case B transition also occurs in any collisionally ionized cloud, as discussed in the first two
papers of this series \citep{2020ApJ...901...68C,2020ApJ...901...69C}. \citet{2020ApJ...901...69C} showed that  the Fe XXV K$\alpha$ line ratios calculated
with Cloudy are in excellent agreement with the line ratios observed by
\textit{Hitomi} for the outer region of Perseus core (see figure 14 in their paper). At the best-fit hydrogen column density of the hot gas at Perseus core ($N_{\rm H, hot}$ = 1.88 $\times$ 10$^{21}$ cm$^{-2}$) reported by \citet{2018PASJ...70...12H}, line formation processes can be best described by Case A.

In extragalactic environments, such as a cloud photoionized by an Active galactic nucleus (AGN) SED with no Lyman absorption lines, the line formation in the low-column-density limit will be described by Case C in the optically thin limit and Case D in the optically thick limit until the cloud becomes very optically thick to stop continuum pumping. 

\begin{figure}[h!]
\centering
\includegraphics[scale=0.5]{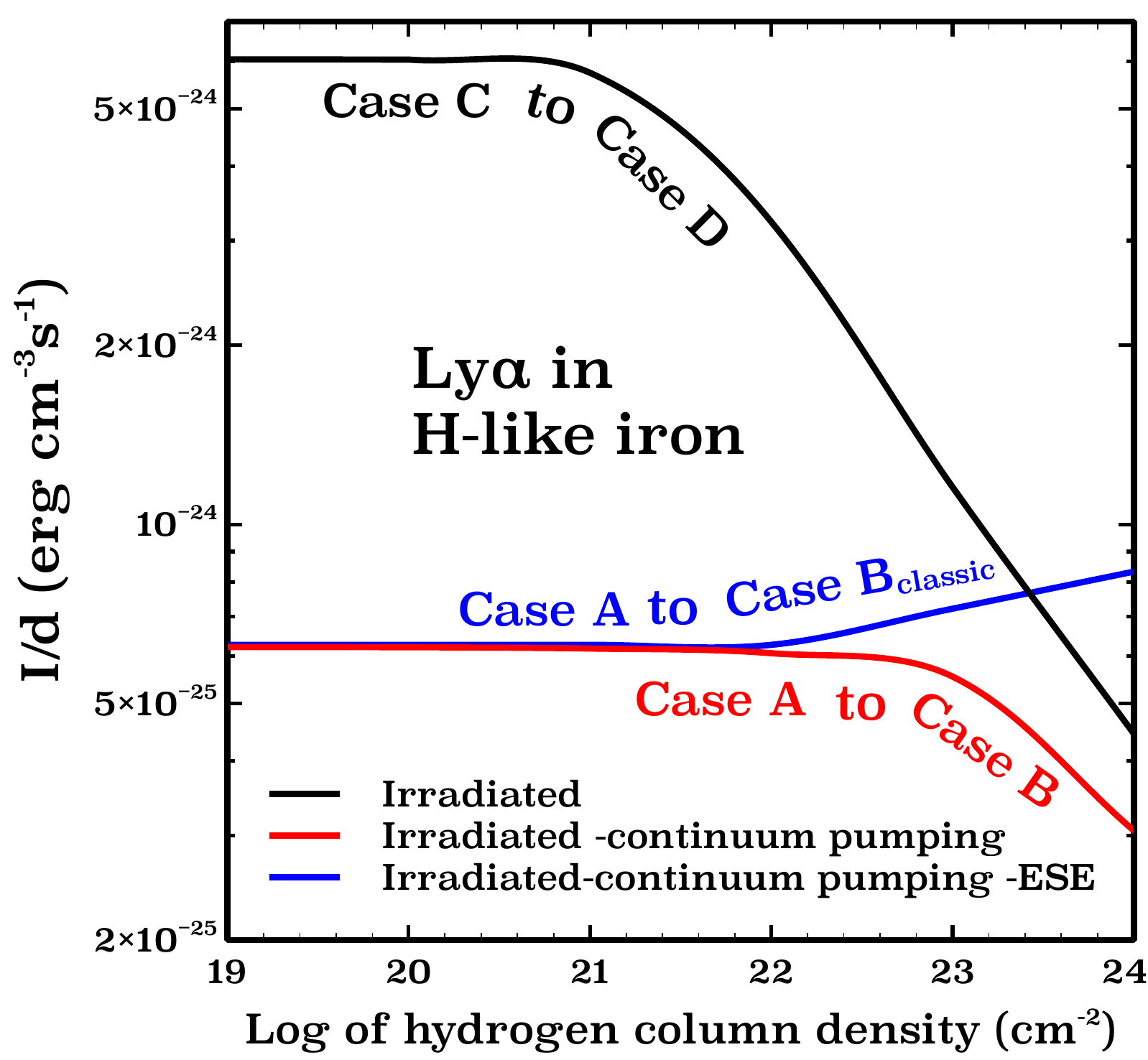}
\caption{Case C to D, Case A to B, and Case A to B$_{classic}$ transitions shown in the same figure for an irradiated cloud in Ly$\alpha$ for H-like iron. The three curves shown in this figure were plotted separately in the left panel of Figure \ref{f:col_den} along with other lines. Case C to D transition curve includes continuum pumping and electron scattering. Case A to B transition curve includes electron scattering but no continuum pumping. Case A to B$_{classic}$ transition curve does not include  continuum pumping or electron scattering, and is the simplest of all three cases.
\label{f:comparison}}
\end{figure}

Figure \ref{f:comparison} shows the variation of I/d in H-like iron with hydrogen column density for the most complex system observed in nature (Case C to D transition) to the simplest possible system ( Case A to B$_{classic}$ transition). Case C to B transition shows the observed I/d values in an irradiated cloud, which includes continuum pumping and electron scattering. Case A to B transition shows the observed I/d with no continuum pumping.  Case A to B$_{classic}$ shows the classical Menzel-Baker transition with no continuum pumping and no electron scattering.

\begin{figure*}
\gridline{\fig{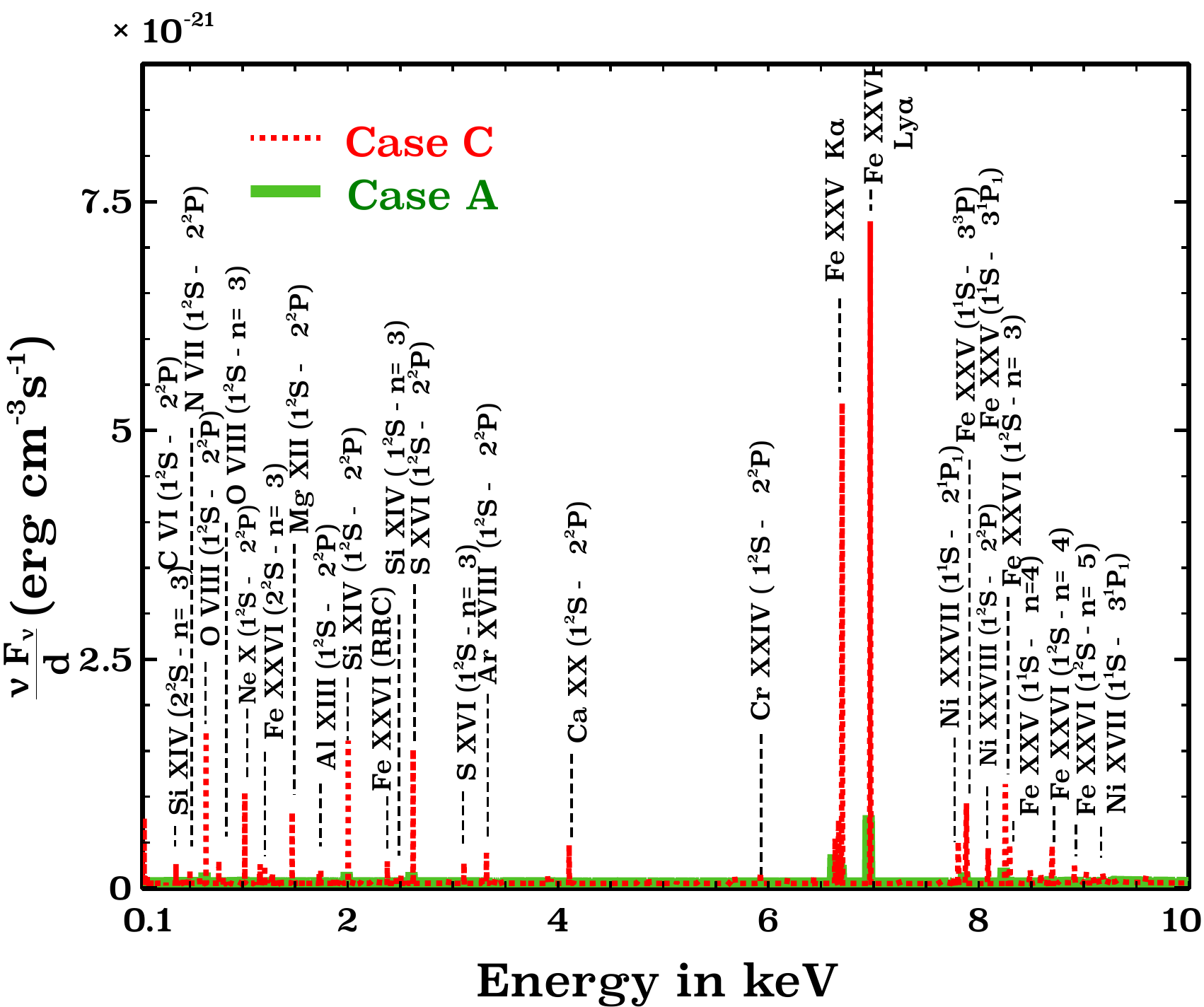}{0.56\textwidth}{(a)}
          \fig{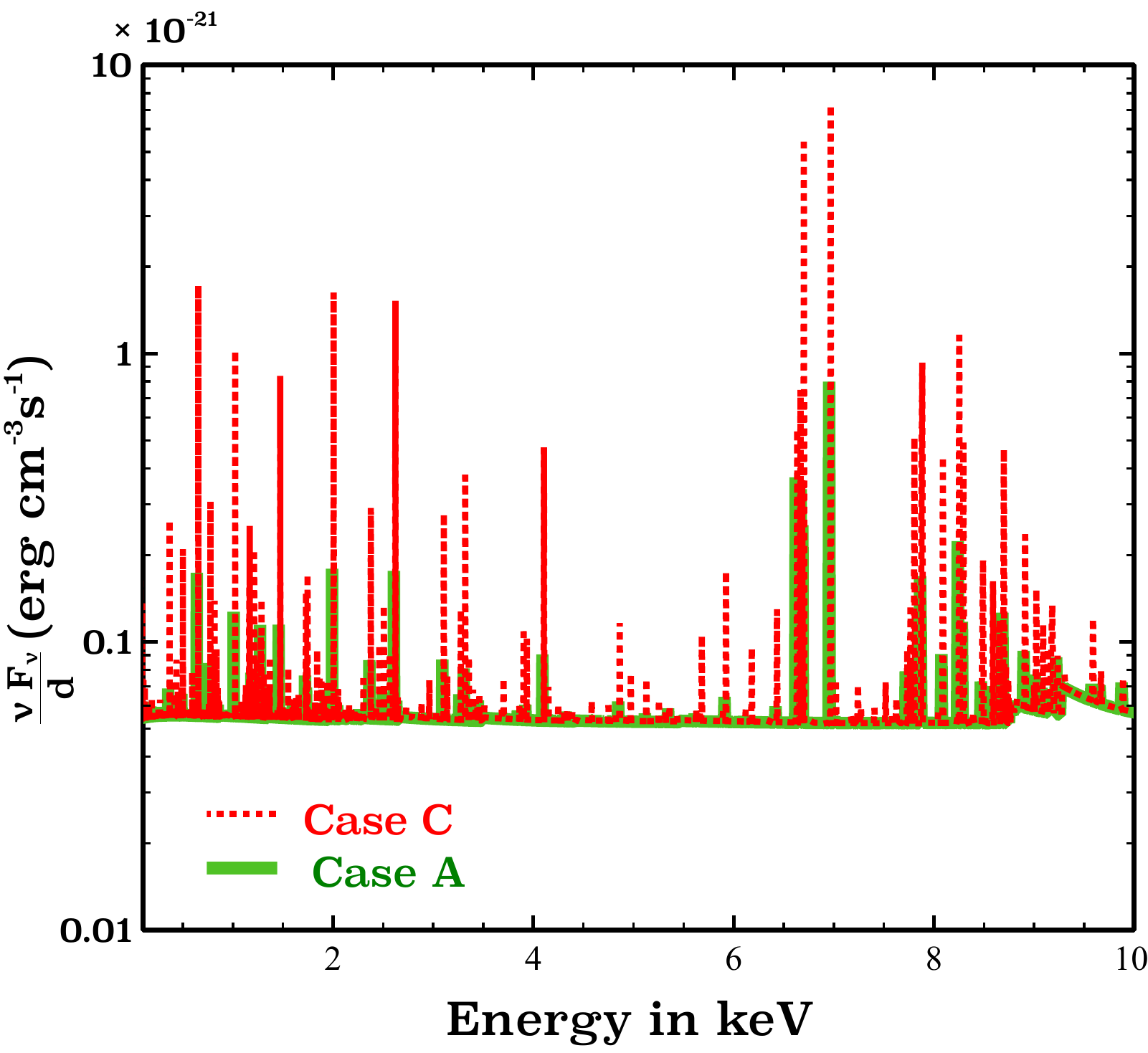}{0.51\textwidth}{(b)} 
          }
\gridline{\fig{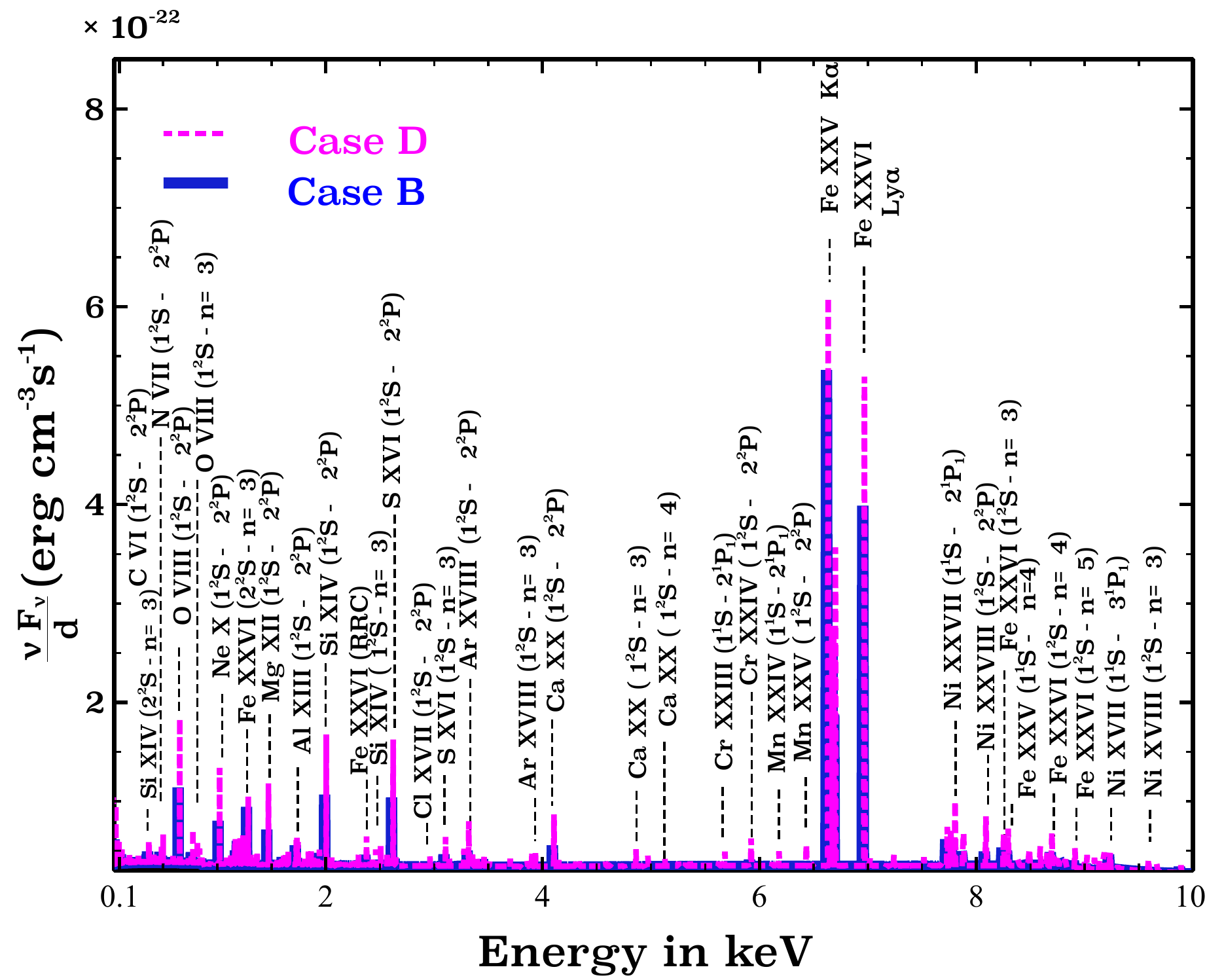}{0.56\textwidth}{(c)}
          \fig{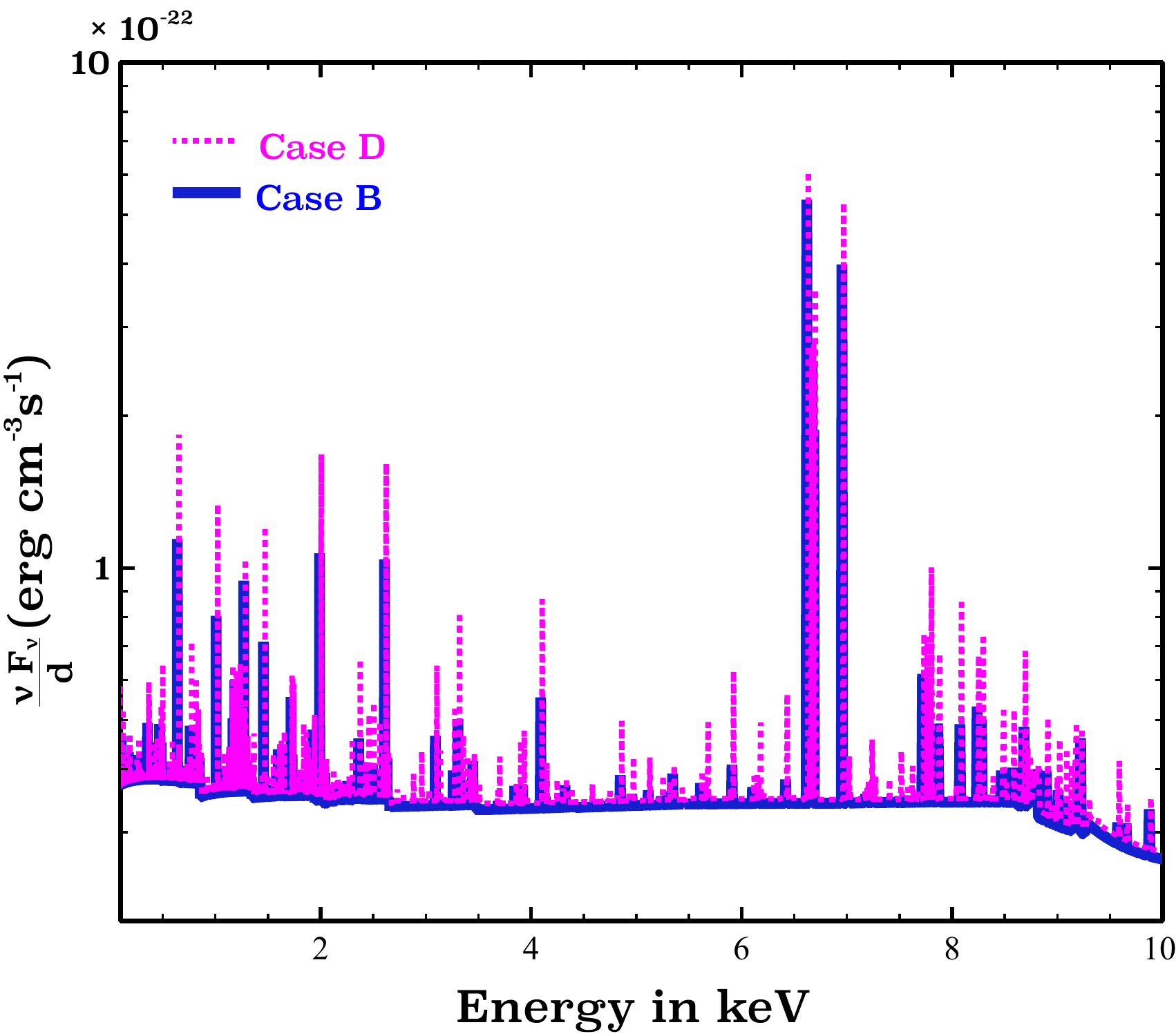}{0.51\textwidth}{(d)}
          }
\caption{ The total scaled emission spectrum for Case A, B, C, and D conditions within the energy range 0.1-10 keV at the resolving power of XRISM (R $\sim$ 1200) around 6 keV. Top-left and top-right panels show Case A (green) and Case C (red) spectrum in linear and log scale. Bottom-left and bottom-right panels show Case B (blue) and Case D (magenta) spectrum in linear and log scale. Case C is brighter than Case A due to continuum pumping, and Case D is brighter than Case B due to partial continuum pumping.
\label{f:3}}
\end{figure*}

\begin{figure}[h!]
\centering
\includegraphics[scale=0.5]{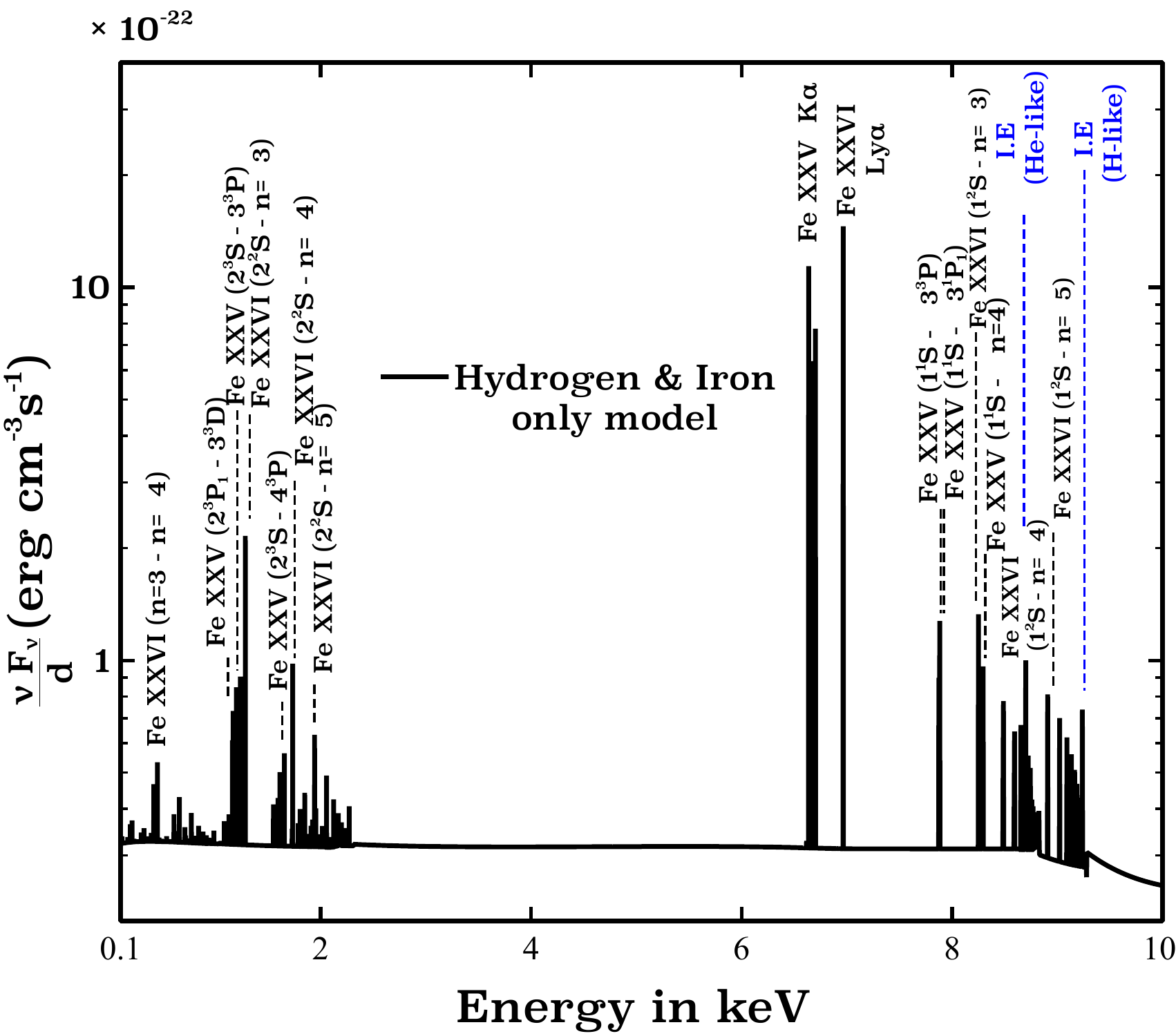}
\caption{A pedagogical simplified Case D spectrum for a hydrogen and iron only model. Lyman and Balmer lines in H- and He- like iron are marked with black dashed lines. Ionization edges (I.E) for H-and He-like iron are shown with blue dashed lines. 
\label{f:Irononly}}
\end{figure}
\section{Description of spectral features}\label{Spectral Features}

Figure \ref{f:3} shows the total observed X-ray spectrum coming from a photoionized cloud  for Case A, B, C, and D within the energy range 0.1-10 keV. The spectrum is generated at the resolving power of XRISM (R $\sim$ 1200) at 6 keV set to Cloudy.
Similar to section \ref{Results}, the y axis of the figure has been scaled to show the total emission ($\nu$ F$_{\nu}$) per unit thickness (d) of the cloud, $\nu$ F$_{\nu}$/d.
$\nu$ F$_{\nu}$ is a sum of the total continuum emission and discrete line intensity (I)
multiplied by R:
\begin{equation}
    \nu F_{\nu} =  \nu F_{\nu}^{continuum} + R I 
\end{equation}

In Cloudy, $\nu$ F$_{\nu}$ can be stored with the following command:
\begin{verbatim}
save emitted continuum 
\end{verbatim}
added to the input script.

The top and bottom rows in  Figure \ref{f:3} overplots the total emission spectrum for
Case A with Case C, and Case B with Case D. The column densities set to 
the cloud for 
calculating the spectra are the same as section \ref{Results}.
Left and right panels
show the same plots linear and log scale, respectively. Clearly, Case C is enhanced 
compared to the Case A spectrum due to continuum pumping. Case D is also brighter
than Case B due to partial continuum pumping. 

Figure \ref{f:Irononly} shows a simplified plot for a hydrogen and iron only model under
Case D condition. This is not what is observed in nature. The purpose of 
this figure is to look at the components of the spectra coming from H- and 
He-like iron in a less complicated form. The Lyman and Balmer
lines are marked with black, and the ionization edges (I.E) of H-like and 
He-like iron at $\sim$ 9.3 keV and $\sim$ 8.8 keV are marked with blue. 


\begin{figure}[h!]
\centering
\includegraphics[scale=0.5]{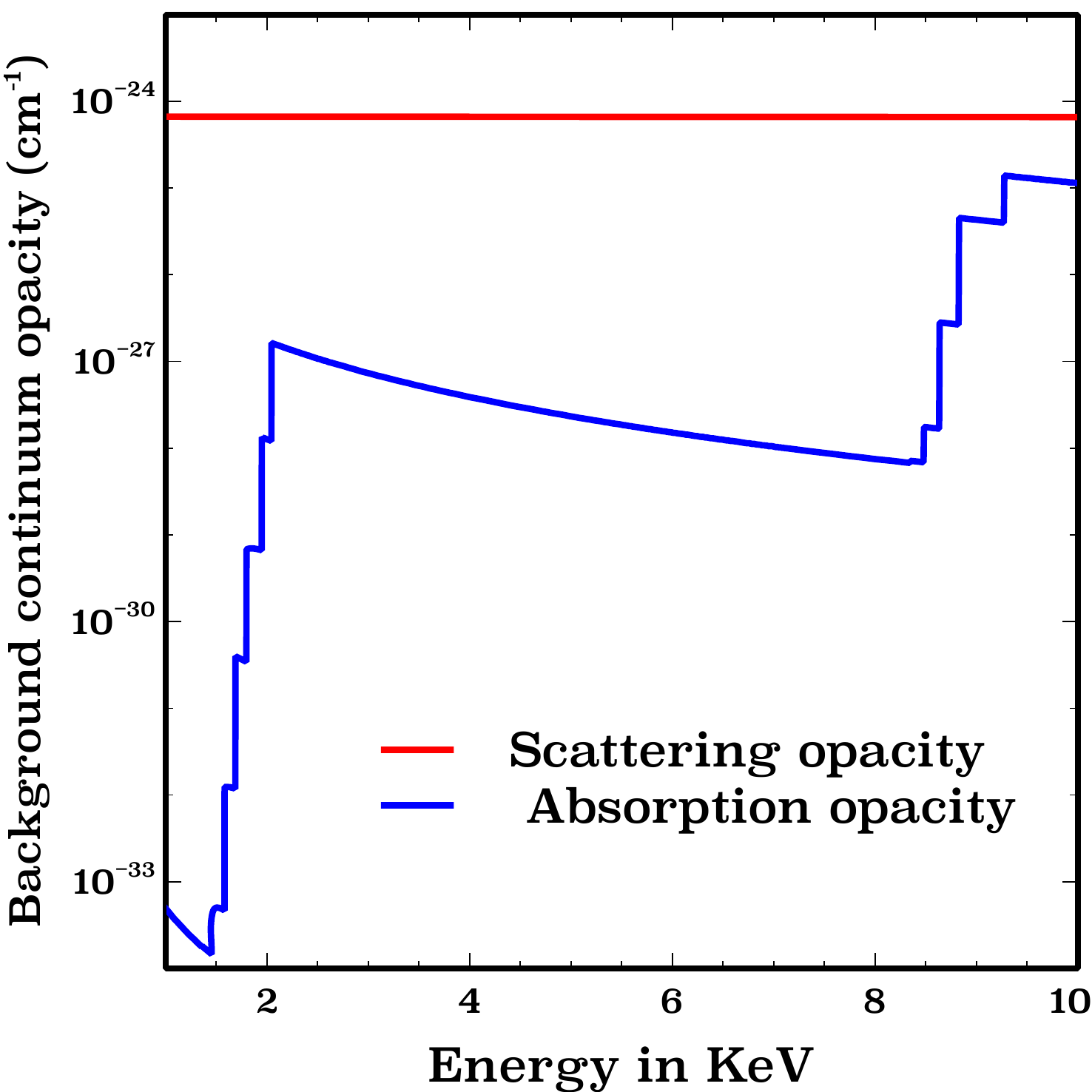}
\caption{Continuum background opacities in a photoionized iron and hydrogen only model. Red solid line shows the total scattering opacity. Blue line shows the total absorption opacity.
\label{f:opacities}}
\end{figure}

\section{Additional factors changing the line intensities}\label{effects of continuum opacities}
The background continuum opacity of the photoionized cloud consists of two types of opacities - absorption opacity and scattering opacity. Figure \ref{f:opacities} shows these two types of opacities. Absorption opacity mostly comes from the photoelectric absorption/photoionization opacity.\footnote{Other absorption opacities like Brems opacity and dust opacity are negligible as Brems opacity depends on the density square, which is very small (1 cm$^{-3}$), and we assume no dust is present in our model.}
 Near the K$\alpha$ complex, absorption opacity is many orders of magnitude smaller than the scattering opacity, making it unimportant without affecting the line spectrum. Thus we only discuss the effects of scattering opacity.


Scattering opacity mostly comes from the scattering of line photons by high-speed thermal electrons that lead to a process called electron scattering escape (ESE). The concept of ESE has been elaborated in section 7 of the first paper of this series \citep{2020ApJ...901...68C}. 

As a result of scattering off high-speed electrons, a fraction of the line photons is
largely Doppler-shifted from their line-center.
These Doppler-shifted photons create super-broad Gaussian profiles. The observed spectrum will include these broad Gaussian profiles as well as the actual sharp line profiles for the fraction of photons that were not scattered \citep{2002ApJ...578..348M, 2009ApJ...690..330H}. 

In Cloudy, these broad Gaussian profiles can be excluded with the following Cloudy command\footnote{ This is a new Cloudy command that counts the intensity of the remaining line photons that didn't suffer electron scattering.
The next update to the release version of Cloudy, C17.03, will include this command.}:
\begin{verbatim}
no scattering intensity 
\end{verbatim}
, reporting only the intensities of the sharp line profiles. 

\begin{figure*}
\centering
\gridline{\fig{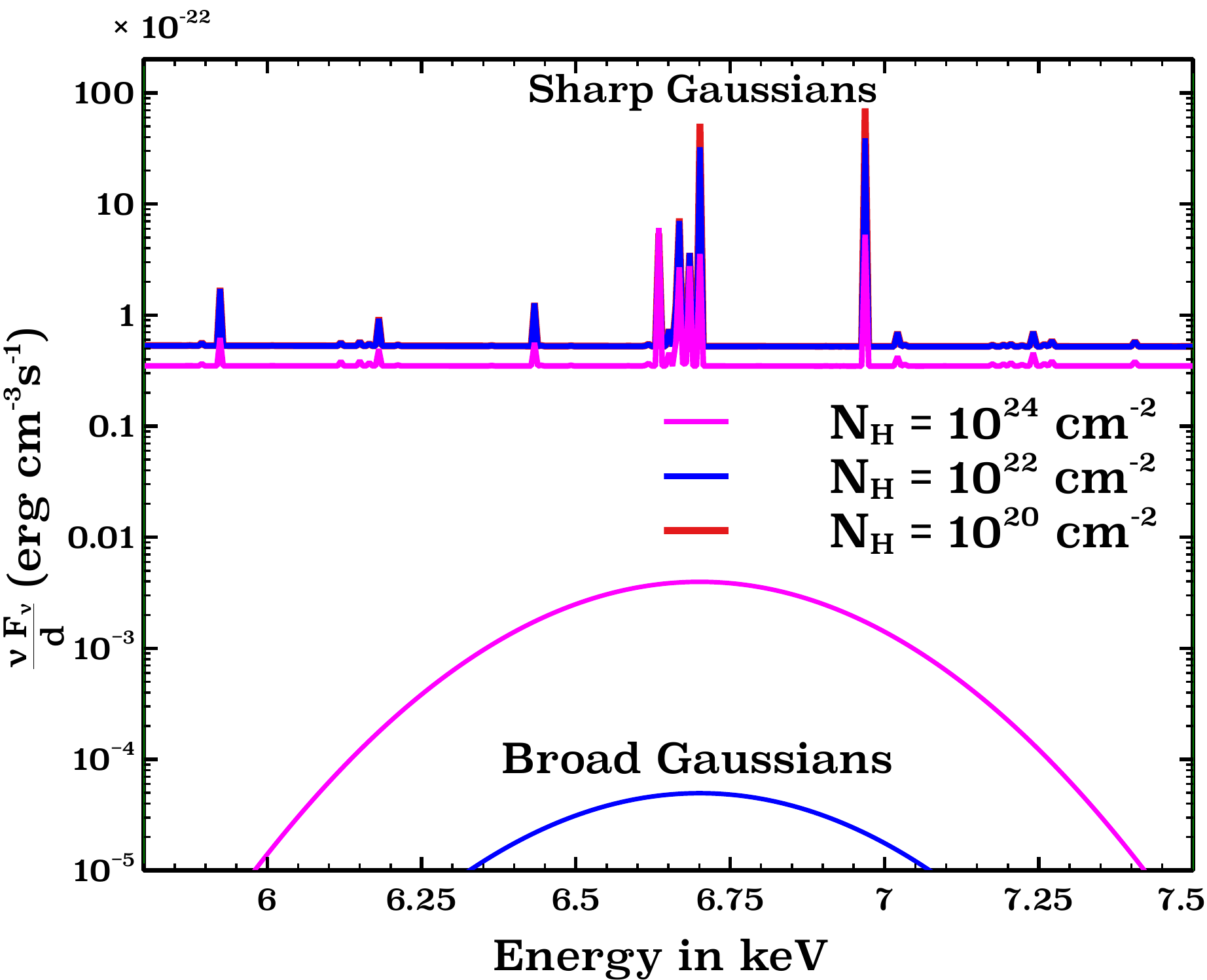}{0.45\textwidth}{(a)}
         \fig{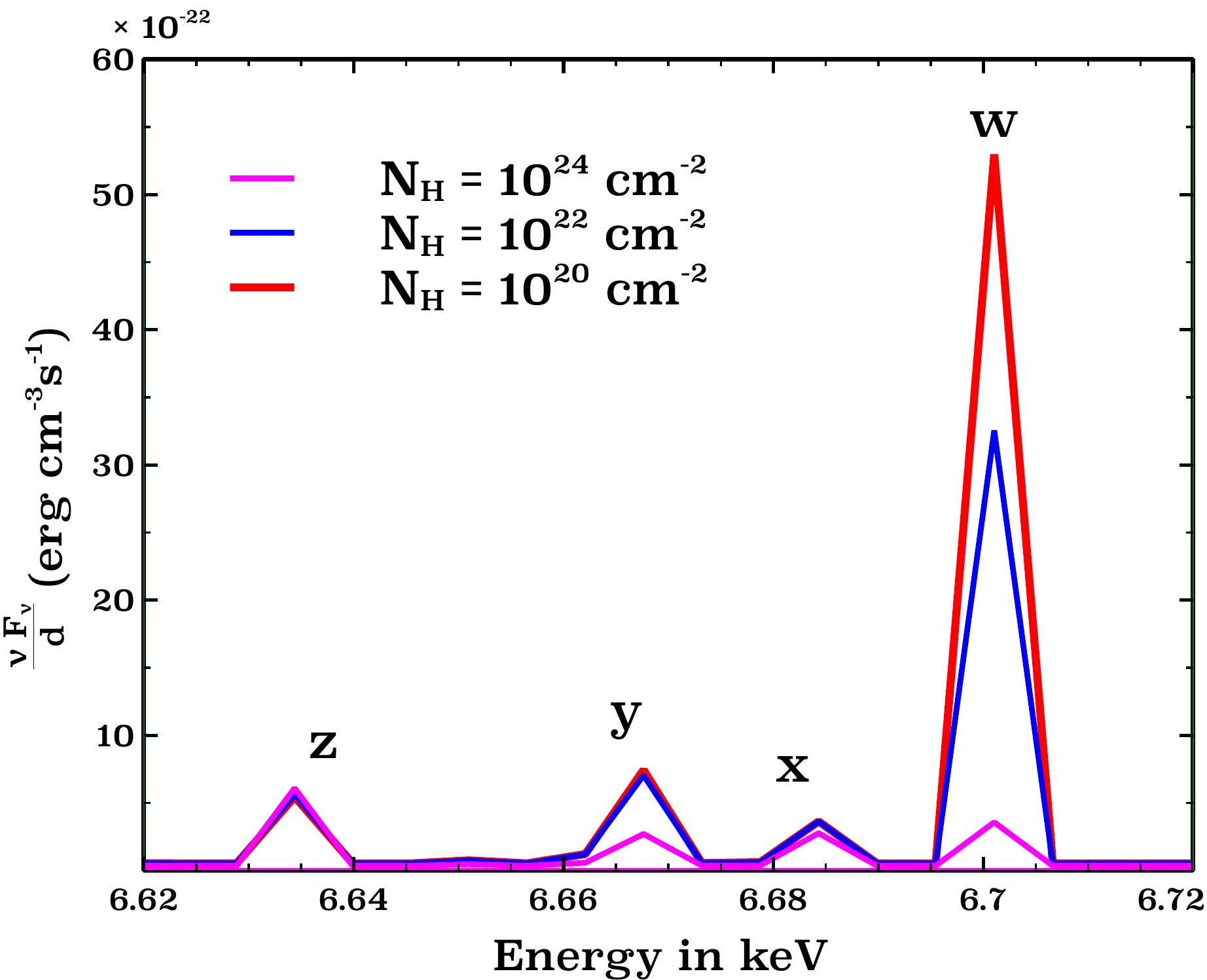}{0.45\textwidth}{(b)} 
         }
\caption{The left panel shows the total scaled emission  (sharp line profiles + continuum including the broad line profiles) in a log scale at
$N_{\rm H}$ =10$^{20}$ cm$^{-2}$, 10$^{22}$ cm$^{-2}$, and 10$^{24}$ cm$^{-2}$ in presence of continuum pumping. The broad Gaussians shown in the figure come
from the photons in the Fe XXV K$\alpha$ complex scattered
off high-speed electrons and Doppler-shifted from their line-centers. The 
sharp line profiles come from the originally emitted line photons in the
photoionized cloud that are not affected by electron scattering. The $N_{\rm H}$ =10$^{24}$ cm$^{-2}$ case has significant optical depth for electron scattering which depresses the continuum.
The right panel is a zoomed-in version of the left panel around the
Fe XXV K$\alpha$ complex. The emission in w significantly reduces with increase in column density. The emission in y reduces slightly. 
\label{f:ese}}
\end{figure*}

Figure \ref{f:ese} shows the total scaled emission near the Fe XXV K$\alpha$ complex. For simplicity, widths of the sharp line profiles are assumed to be coming from the thermal velocity of the iron ions only. The presence of turbulence will change the widths of the sharp components, but the physics of electron scattering will be the same.
At T = 6 $\times$ 10$^{6}$ K, the temperature of our
simulated cloud, the FWHM of these sharp line profiles at E $\sim$ 6.7 keV are: $\Delta$ $E_{\rm FWHM}^{\rm sharp}$ = 2 $\sqrt{ln2}$ $\frac{u_{\rm Dop}}{c}$ E $\sim$ 1.6 eV, where $u_{\rm Dop}$= $\sqrt{\frac{2kT}{m_{\rm Fe}}}$ $\sim$ 43 km/s. At the same temperature, FWHM of the broad line profiles are $\Delta$ $E_{\rm FWHM}^{\rm broad} \sim$ 0.5 keV, where $u_{\rm Dop}$= $\sqrt{\frac{2kT}{m_{\rm e}}}$ $\sim$ 13500 km/s . This implies that, the height of  broad Gaussians will be orders of magnitude smaller than the sharp components, and are difficult to detect by telescopes \citep{2010ApJ...715..947T}. 


 The left panel in Figure \ref{f:ese} shows the changes in the 
 total scaled emission in a log scale for the hydrogen column densities $N_{H}$ =10$^{20}$ cm$^{-2}$, 10$^{22}$ cm$^{-2}$, and 10$^{24}$ cm$^{-2}$ in the presence of continuum pumping. The broad Gaussians shown in the figure are solely from the
 electron scattering of the photons in Fe XXV K$\alpha$ complex. We do not show the 
 broad Gaussians from the other lines to keep the figure simple.
 
 The right panel shows a zoomed-in version of the sharp line profiles for all three column densities on a linear scale.  As the continuum pumping is present, $\nu$ F$_{\nu}$/d at $N_{H}$ =10$^{20}$ cm$^{-2}$ represents the Case C limit, and $N_{H}$ =10$^{24}$ cm$^{-2}$ represents the Case D limit. 

It can be seen from the figure that w line intensity reduces significantly with the increase in optical depth/column density. There are two factors responsible for this reduction.
First, the continuum pumping begins to become partially blocked with the
increase in optical depth.
 Second, w has the largest line-optical depth among all He-like transitions and  is more likely to suffer
electron scattering. Of course, when observed by future high-resolution  telescopes like XRISM and Athena, the electron scattered broad Gaussian component of w will be much fainter than the sharp component. But the reduction in the sharp w line intensity
(or line intensity of any resonance line) with increasing optical depth 
can serve as a  powerful optical depth/column density diagnostic.

Note that, \citet{1987SvAL...13....3G} discussed the effects of resonance scattering and showed the distortion in radial 
surface brightness profile due to migration of resonance X-ray line photons from the cluster center to the outer region. This will lead to a suppression in line fluxes in the central region of a cluster.
For example, in Perseus, this factor was reported to be $\sim$ 1.3 for w ($\tau$ $\sim$ 1) near the cluster
center by \citet{2018PASJ...70...10H}. Of course, the suppression factor will be 
different in other systems depending on the geometry and optical depth.

Our calculations presented in this paper represent a general study for a 
photoionized system irradiated with a power-law SED. From our Cloudy calculation,
the suppression in w line intensity at
$\tau$ = 1 is $\sim$ 1.43 due to the joint contribution of
partial continuum
pumping and electron scattering.
Therefore it is safe to say that, the change in the resonance line intensities 
due to these two factors can be as important as the resonance scattering 
effects. Our current model predicts the total emission from a symmetric geometry, so scattering has no effect on the emergent intensity. 
In addition to what we report in this paper, the \citet{1987SvAL...13....3G} resonance scattering geometric correction
has to be applied to match with the observed spectra.


\section{Discussion and Conclusion}\label{summary}

\begin{itemize}

\item
Line formation processes were broadly categorized into two cases in the 1930s -  Case A and Case B \citep{1937ApJ....85..330M, 1937ApJ....86...70M, 1938ApJ....88...52B, 1938ApJ....88..422B}. At that time, the SED ionizing the cloud was assumed to have strong Lyman absorption lines. There was no continuum pumping/fluorescence to enhance the spectra. Some examples are O-stars in starburst galaxies, planetary nebulae etc \citep{2006agna.book.....O}. 

But in extragalactic environments such  as  a  cloud  photoionized by an AGN SED, or galaxies with quasars with no Lyman absorption lines, continuum pumping will significantly enhance the spectra. This lead to the discovery of a third case in the late 1930 s - Case C \citep{1938ApJ....88..422B}, which describes optically thin clouds in the presence of continuum pumping. 
A fourth case - Case D was discovered recently \citep{2009ApJ...691.1712L}, which describes spectra from an optically thick cloud in the presence of continuum pumping. 
\end{itemize}

\begin{itemize}
\item
Figure \ref{f:1} shows the schematic representation of all four cases.  
     \begin{itemize}
     \item Under Case A condition, lines are formed by radiative recombination and cascades from higher levels. Lyman lines escape the optically thin cloud without scattering/absorption. 
     \end{itemize}
    
    \begin{itemize}
     \item Case B occurs when higher-n Lyman lines are converted to Balmer lines and  Ly$\alpha$ (or K$\alpha$) or two-photon-continuum due to multiple scatterings in an optically thick cloud.
     \end{itemize}
    
    \begin{itemize}
     \item Lines are formed under Case C when Lyman lines are enhanced by continuum pumping and freely escape the optically thin cloud.
      \end{itemize}
     
\begin{itemize}
     \item Case D occurs when multiple scattering and continuum pumping in Lyman
     lines occur together in an optically thick cloud so that the lines
     are partially
     enhanced.
     \end{itemize}
\end{itemize}

\begin{itemize}
\item
This paper is dedicated to understanding line formation processes through Case A, Case B, Case C, and Case D in the X-ray emitting photoionized plasma with Cloudy. We study H- and He-like iron emitting
in the X-ray with improved Cloudy energies in excellent agreement with the future microcalorimeter observations \citep{2020RNAAS...4..184C}. In our simulations, we use a power-law SED to illuminate the cloud, and an equilibrium temperature of 6 $\times$ 10$^{6}$ K  computed from the heating-cooling balance. Refer to section \ref{Simulation Parameter} for details on simulation parameters. As the absolute line intensities (I) increase with cloud thickness (d), we compare
 line intensity per unit thickness of the cloud (I/d) for estimating the scaled difference between all four cases.

\end{itemize}

\begin{itemize}
     \item 
     Table \ref{t:1} lists the line intensity (I) per unit thickness (d) of the cloud, I/d, for the Case A, Case B, Case C, and Case D conditions observed in nature. 
      To generate the optically thin and optically thick conditions in the cloud, I/d's for Case A and C are computed at N$_{H}$=10$^{19}$ cm$^{-2}$, and for Case B and D  are computed at N$_{H}$=10$^{24}$ cm$^{-2}$, respectively.
      The Menzel-Baker Case B values are listed under the  Case B$_{classic}$  column in the table.
      Ly$\alpha$ in H-like iron and K$\alpha$ in He-like iron in Case B$_{classic}$ are enhanced compared to their corresponding Case A values due to the conversion of higher n-Lyman lines into Ly$\alpha$ (or K$\alpha$) plus Balmer lines.
      But in real astronomical sources, the presence of electron scattering reduces the observed Case B values. In H-like iron, I/d for Ly$\alpha$ decreases by $\sim$ 50\%. In He-like iron, x, y, and w exhibit a decrease up to $\sim$ 24 \%.
      Case C values are the brightest of all four cases due to the free escape of Lyman photons following continuum pumping. The Ly$\alpha$ and K$\alpha$ transitions in H- and He-like iron are up to $\sim$ 10 and $\sim$ 27 times enhanced compared to the corresponding Case A values. Case D values are smaller than Case C values but bigger than the Case B values, as they are partially enhanced by continuum pumping. For H like iron, Case D I/d for Ly$\alpha$ is $\sim$ 36\% enhanced compared to the corresponding Case B value. Ly$\beta$ is $\sim$ 71\% enhanced, H$\alpha$ is $\sim$ 36\% enhanced.
      In He-like iron, K$\alpha$ is enhanced up to $\sim$ 109\%, K$\beta$ is enhanced up to $\sim$  208\%, L$\alpha$ is enhanced up to $\sim$ 80\%.
 \end{itemize}    
     
   \begin{itemize}
     \item   
      The total emission spectrum for Case A, B, C, and D conditions within
      the energy range 0.1-10 keV have been shown in Figure \ref{f:3}. The spectrum
      includes the continuum emission as well as the line emission described in the previous paragraphs. The 
      resolving power (R) for our Cloudy simulations is set at R $\sim$ 1200, 
      which is the resolving power of XRISM at $\sim$ 6 keV. The 
      figure shows Case A overplotted with Case C and Case B overplotted 
      with Case D in linear and log scale. Clearly, the line emissions 
      in the Case C spectrum are brighter than Case A, and line emissions in the Case D spectrum are brighter than Case B due to continuum pumping and partial continuum pumping, respectively.
     
   \end{itemize}


 \begin{itemize}
\item Electron scattering opacity can play an important role in deciding the line intensities in optically thick clouds. Line intensities for Case B and Case D can be reduced because of this. The top panel of Figure \ref{f:col_den} and Table \ref{t:1} showed the deviation of the observed Case B values, which includes the effect of electron scattering, from Case B$_{classic}$, which does not include electron scattering. I/d values shown in Table \ref{t:1} and the bottom panel of
Figure \ref{f:col_den} for  Case D also include electron scattering.
\end{itemize}

\begin{itemize}
\item
Due to the electron scattering opacity, the line photons are scattered off high-speed electrons and are Doppler shifted from their line-center. These scattered photons form Gaussians with super-broad-bases. The line photons that are not scattered have a sharp base equivalent to their thermal width (and turbulent width if turbulence is present).
Figure \ref{f:ese} shows these sharp and broad components coming from Fe XXV K$\alpha$ 
complex in the presence of continuum pumping for the hydrogen column densities $N_{H}$ =10$^{20}$ cm$^{-2}$, 10$^{22}$ cm$^{-2}$, and 10$^{24}$ cm$^{-2}$.
$N_{H}$ = 10$^{20}$ cm$^{-2}$ is the Case C limit, and $N_{H}$ = 10$^{24}$ cm$^{-2}$ is the Case D limit.
The broad components will be much fainter than the sharp  components when detected by high-resolution telescopes. The observed sharp components for the resonance lines will exhibit significant changes in their line-fluxes with variation in column density (and optical depth). It can be seen in Figure \ref{f:ese} that, the w line-intensity decreases significantly with an increase in column density. Such reduction in w is due to the two following factors - a) Continuum pumping becomes partially blocked with the increase  in  optical  depth. b) The large optical  depth of w makes it more likely to be scattered by electrons. 
A combination of a) \& b) will reduce the line intensity of w or
any resonance line significantly with the increase in column density, which can serve as a powerful diagnostic in measuring the column density/optical depth of the cloud. 

From our Cloudy simulation, we get the suppression in  w line intensity due 
to the two above factors at $\tau$=  1 to be $\sim$ 1.43, which is as important as the
resonance scattering effects described by \citet{1987SvAL...13....3G}. As 
Our current Cloudy model assumes
a  symmetric geometry, the effects of resonance scattering are not
included in our calculation. A real observed spectrum will correspond to a resonance-scattering corrected Cloudy-generated spectrum shown in 
this paper.

\end{itemize}

\begin{itemize}
 
 \item 
 After the discovery of Case A and B, these two cases have been widely discussed in the literature for the optical, UV, and infrared regimes, with limited studies on X-rays. As far as we know, there has been no discussion on X-ray spectra under Case C and Case D condition. 
 Case D is the least discussed of all four cases, as ideally, at very high column densities, 
 Case D should be no different than Case B values, as mentioned in section \ref{Case D}. 
But Table \ref{t:1} and bottom panel of Figure \ref{f:3} shows that, even at a column density as high as N$_{H}$=10$^{24}$  cm$^{-2}$ in a cloud illuminated with a power-law SED, Case D deviates considerably from Case B for X-ray emission from H- and He-like iron. This deviation will certainly be detected by the future high-resolution telescopes with microcalorimeter technology. \\
We emphasize the fact that Case C and Case D deserve far more 
attention than they have  been given to date, especially because
they could be the best representation of the emission spectra from irradiated
extragalactic sources with a broad range of column densities.
\end{itemize}

\section*{Acknowledgement}
We acknowledge the referee of this paper for his/her very helpful comments. We thank Stefano Bianchi and Anna Ogorzalek for their valuable comments.
We acknowledge support by NSF (1816537, 1910687), NASA (17-ATP17-0141, 19-ATP19-0188), and STScI (HST-AR-15018).
MC also acknowledges support from STScI (HST-AR-14556.001-A).


\bibliography{references}{}
\bibliographystyle{aasjournal}

\end{document}